\documentclass[11pt,a4paper]{article} 
\usepackage{a4wide}
\usepackage{amsmath}
\usepackage{graphicx}
\usepackage{makecell}
\usepackage{verbatim}
\usepackage{multirow}
\usepackage{subfig}
\usepackage{bm}
\usepackage{braket}
\usepackage{mathtools}
\usepackage{amssymb}
\usepackage{amscd}
\usepackage{latexsym}
\usepackage{slashed}
\usepackage[normalem]{ulem}
\usepackage{verbatim}
\usepackage{rotating}
\usepackage{diagbox}
\usepackage{cancel}
\usepackage[utf8]{inputenc}
\usepackage{array}
\usepackage{epsfig,amsfonts,amsthm}
\usepackage{float}
\newlength{\absize}
\usepackage{verbatim}
\usepackage{cite}
\numberwithin{equation}{section}

\begin{document}

\thispagestyle{empty}
\renewcommand{\thefootnote}{\fnsymbol{footnote}}
\newpage\normalsize
\pagestyle{plain}
\setlength{\baselineskip}{4ex}\par
\setcounter{footnote}{0}
\renewcommand{\thefootnote}{\arabic{footnote}}
\renewcommand{\title}[1]{
\begin{center}
\LARGE #1
\end{center}\par}
\renewcommand{\author}[1]{
\vspace{2ex}
{\Large
\begin{center}
 \setlength{\baselineskip}{3ex} #1 \par
\end{center}}}
\renewcommand{\thanks}[1]{\footnote{#1}}
\renewcommand{\abstract}[1]{
\vspace{2ex}
\normalsize
\begin{center}
\centerline{\bf Abstract}\par
\vspace{2ex}
\parbox{\absize}{#1\setlength{\baselineskip}{3.5ex}\par}
\end{center}}

\vspace*{4mm}

\title{\textbf{Spontaneous CP violation in the \boldmath{$D_5$}-symmetric four‑Higgs‑doublet models} }

\author{Dong-Ping Fu$^{a}$\footnote{E-mail: fudongp@mail2.sysu.edu.cn} 
Michihisa Takeuchi$^{a}$\footnote{E-mail: takeuchi@mail.sysu.edu.cn}}

\begin{center}
$^{a}$School of Physics and Astronomy, Sun Yat-sen University, 519082 Zhuhai, China 
\end{center}

\vfill

\abstract{ We constructed a four-Higgs-doublet model (4HDM) invariant under $D_5$ symmetry and investigated its complete neutral vacuum structure in detail. Assuming explicit CP conservation in the scalar potential, we examined whether CP symmetry can be spontaneously broken. We provided a complete list of all possible real and complex vacua, along with the constraints on the potential parameters required for each vacuum solution to exist. We also discussed the positive-definiteness conditions that the Hessian must satisfy for each vacuum to be a local minimum of the potential. The results show that, after spontaneous symmetry breaking, some complex vacua lead to spontaneous CP violation in the potential, whereas the remaining complex vacua still preserve CP conservation. Among these CP-violating complex vacua, one can be regarded as the most general form. Furthermore, we discussed the relationship between real and complex vacua. }
\vspace*{30mm} \setcounter{footnote}{0} \vfill
\newpage
\renewcommand{\thefootnote}{\arabic{footnote}}

\setcounter{page}{1}
\section{Introduction}
\label{Sec:intro}
With the 2012 announcement by the Large Hadron Collider (LHC) \cite{ATLAS:2012yve,CMS:2012qbp} of a Higgs boson with a mass of approximately 125 GeV, the Standard Model (SM) of particle physics received decisive experimental confirmation. Owing to these numerous experimental confirmations, the SM has become the most successful theory for describing the behavior and interactions of elementary particles. However, many puzzles remain beyond the explanatory power of the SM. These include, for example, that the strength of CP violation in the SM is insufficient to generate the matter-antimatter asymmetry of the universe \cite{Sakharov:1967dj,Kuzmin:1985mm}, and that neutrino oscillation experiments have established that neutrinos have tiny masses \cite{Super-Kamiokande:1998kpq}, whereas in the SM neutrinos are strictly massless.

In the SM, there is only one Higgs doublet. For a long time, there has been no strong experimental evidence indicating that only one type of Higgs boson exists in nature. As a result, extensions of the SM Higgs sector—namely, multi-Higgs-doublet models (MHDMs)—have emerged as an important research direction in studying new physics beyond the SM \cite{Keller:1986nq,Casalbuoni:1987cz,Grossman:1994jb,Asakawa:1999gz,Grimus:2002nk,Barroso:2006pa,Grimus:2007if,Grimus:2008nb,Ferreira:2010xe,Pilaftsis:2016erj,Bento:2017eti,Darvishi:2019dbh}. One important motivation for studying MHDMs is that explicit or spontaneous CP violation in their scalar potentials can provide additional sources of CP violation \cite{Branco:1985pf,Wu:1994ja,Lavoura:1994fv,Branco:2005em,Nishi:2006tg,Inoue:2014nva,Barradas-Guevara:2015rea,deMedeirosVarzielas:2016rii,Grzadkowski:2016szj,Emmanuel-Costa:2016vej,Nierste:2019fbx,Nebot:2019qvr,Okada:2020brs,Kuncinas:2023ycz,Miro:2024zka}, thereby helping to address the matter-antimatter asymmetry of the universe. The study of MHDMs traces back to the two-Higgs-doublet model (2HDM) proposed by T. D. Lee in 1973 \cite{Lee:1973iz}, in which spontaneous CP violation (SCPV) originating from the relative phase of the vacuum expectation values (vevs) provides an additional source of CP violation. Subsequently, in 1976, S. Weinberg introduced the three-Higgs-doublet model (3HDM), providing a natural theoretical explanation for the milliweak origin of CP violation \cite{Weinberg:1976hu}. To date, the 2HDM and 3HDM have been extensively studied, with numerous related results continuing to emerge \cite{Deshpande:1977rw,Pakvasa:1977in,Kanemura:1999xf,Davidson:2005cw,Eriksson:2009ws,Mahmoudi:2009zx,Machado:2010uc,Branco:2011iw,Crivellin:2013wna,Keus:2013hya,BhupalDev:2014bir,Maniatis:2014oza,Akeroyd:2016ssd,Misiak:2017bgg,Das:2019yad,Hernandez:2021iss,Darvishi:2021txa,Bento:2022vsb}. However, studies of the four-Higgs-doublet model (4HDM) remain relatively scarce \cite{Shao:2023oxt}. 

In existing 4HDM studies, such models are typically employed to explain the hierarchical structure of the masses and the CKM mixing matrix for the three generations of fermions, as well as CP violation \cite{Rajpoot:1988gw,Srivastava:2021pyz,Gao:2024xte}. They have also been used to account for dark matter within supersymmetric extensions of the SM \cite{Kawase:2011az}, and to construct new neutrino models in grand unified theories \cite{CarcamoHernandez:2022vjk}. A primary reason for the limited exploration of the 4HDM is that MHDMs typically involve numerous free parameters. Compared with the 3HDM, the additional Higgs doublet in the 4HDM further increases the number of free parameters, which in turn renders the analysis of the vacuum structure, phenomenology, and other aspects of the model more complex and challenging.

By introducing appropriate discrete symmetries, the number of free parameters in a model can be effectively reduced. In this work, we impose the $D_5$ symmetry group on the 4HDM. Compared with other symmetries, the advantage of $D_5$ lies in its ability to significantly reduce the number of free parameters in the 4HDM, leaving only 16 real parameters. This number is comparable to that of certain 3HDMs (e.g., the $Z_2$ 3HDM \cite{Keus:2013hya}, which also has 16 real parameters), thereby substantially reducing the complexity of studying the model. 

Previous studies employing $D_5$ symmetry have often aimed to explain fermion masses and mixing \cite{Krawczyk:1980bb,Ma:2004br,Hagedorn:2006ir,Vien:2019eju,Van:2023aib,Pramanick:2023fal}, with one of the earliest applications being a four-quark model \cite{Krawczyk:1980bb}, in which three Higgs doublets were introduced. Later, C. Hagedorn et al. \cite{Hagedorn:2006ir} attempted to implement $D_5$ symmetry in the 3HDM, but found that its scalar potential exhibits an accidental $U(1)$ symmetry that would lead to phenomenologically unacceptable massless Goldstone bosons. Building on this, they successfully introduced $D_5$ symmetry into the 4HDM and proved that the scalar potential of this model has no accidental symmetries. However, their work only demonstrated that spontaneous CP violation is not allowed for completely arbitrary vev parameters; they neither investigated the specific vev configurations that might lead to SCPV nor analyzed the full vacuum structure of the model. In general, a thorough study of the full vacuum structure is crucial for any successful MHDM. In this paper, we investigate the full neutral vacuum structure of the 4HDM with $D_5$ symmetry. We systematically analyze all possible real and complex vacua, focusing on the possibility of spontaneous CP breaking. Finally, we provide a complete list of the complex vacua that lead to spontaneous CP violation.

The remainder of this paper is organized as follows. Section~\ref{Sec:potential} introduces the non-Abelian discrete group $D_5$ and incorporates it into the 4HDM, thereby constructing the scalar potential of the model. Section~\ref{Sec:vacua} systematically summarizes the procedure for solving the neutral vacuum structure. Section~\ref{Sec:Realvacua} analyzes the possible real vacua in the $D_5$ 4HDM and presents the conditions for the positive definiteness of the Hessian. Section~\ref{Sec:Complexvacua} analyzes the possible complex vacua in the $D_5$ 4HDM, investigates whether they can lead to spontaneous CP violation, and discusses the corresponding positive-definiteness conditions. Section~\ref{Sec:Z2} discusses the residual $Z_2$ symmetry. Section~\ref{Sec:Complexvsreal} discusses the relationship between real and complex vacua. Finally, Section~\ref{Sec:Conclusion} summarizes our conclusions. 

\section{The $D_5$ symmetry and the scalar potential}
\label{Sec:potential}
\subsection{Representation and tensor product}

The dihedral group $D_5$ is the symmetry group of a regular pentagon. It is isomorphic to $Z_5 \rtimes Z_2$ and has ten elements, denoted by $a^{m}b^{k}$ with $m = 0, 1, 2, 3, 4$ and $k = 0, 1$, where the generators $a$ and $b$ satisfy $a^5 = e$, $b^2 = e$, and $bab = a^{-1}$. $D_5$ has four irreducible representations: two singlets, $\mathbf{1}$ and $\mathbf{1}'$, and two doublets, $\mathbf{2}$ and $\mathbf{2}'$. The generators for the singlets of $D_5$ are given in~\cite{Ishimori:2010au}.
\begin{equation}
\label{D5g1}
\begin{aligned}
&\mathbf{1}: \quad a=1, \quad b=1,\\&
\mathbf{1}': \quad a=1, \quad b=-1.
\end{aligned}
\end{equation}
We next consider the complex representation of $D_5$. We choose a complex basis in which the generators of the doublet representation of $D_5$ are given by~\cite{Ishimori:2010au}
\begin{equation}
\label{D5Cg2}
\begin{aligned}
&\mathbf{2} : \quad
a=\begin{pmatrix}
e^{\frac{2i\pi}{5}} & 0 \\
0 &e^{-\frac{2i\pi}{5}}
\end{pmatrix},
\quad b=\begin{pmatrix}
0 & 1 \\
1 & 0
\end{pmatrix},\\&
\mathbf{2}' : \quad
a=\begin{pmatrix}
e^{\frac{4i\pi}{5}} & 0 \\
0 &e^{-\frac{4i\pi}{5}}
\end{pmatrix},
\quad b=\begin{pmatrix}
0 & 1 \\
1 & 0
\end{pmatrix}.
\end{aligned}
\end{equation}
When the two doublets, $x=(x_1 , x_2)^\intercal $ and $ y=(y_1 , y_2)^\intercal $, are in the doublet representation of $D_5$, we denote them by:
\begin{equation}
\left( \begin{array}{c}{x_1 } \\ {x_2} \end{array}\right)\sim \mathbf{2}, \qquad 
\left( \begin{array}{c}{y_1 } \\ {y_2} \end{array}\right)\sim \mathbf{2}',
\end{equation}
then
\begin{equation}
\left( \begin{array}{c}{x^\dagger_2 } \\ {x^\dagger_1  }\end{array}\right)\sim \mathbf{2}, \qquad 
\left( \begin{array}{c}{y^\dagger_2 } \\ {y^\dagger_1  } \end{array}\right)\sim \mathbf{2}'.
\end{equation}

In $D_5$, in the basis of Eqs.~(\ref{D5g1}) and (\ref{D5Cg2}), the Clebsch-Gordan coefficients for all tensor products are as follows. The tensor product of the singlet $w$ with the singlet $z$ yields
\begin{equation}
\label{D511}
\begin{aligned}
&(w)_{\mathbf{1}}\otimes  (z)_{\mathbf{1}}=(w)_{\mathbf{1}'}\otimes (z)_{\mathbf{1}'} =(wz)_{\mathbf{1}}, \qquad (w)_{\mathbf{1}}\otimes  (z)_{\mathbf{1}'}=(wz)_{\mathbf{1}'}.
\end{aligned}
\end{equation}
The singlet $w$, together with the doublet $x=(x_1, x_2)^\intercal$, gives
\begin{equation}
\label{D5C12}
\begin{aligned}
&(w)_{\mathbf{1}}\otimes 
\begin{pmatrix}
x_1 \\
x_2 
\end{pmatrix}_{\bar{\mathbf{2}} }
=\begin{pmatrix}
wx_1 \\
wx_2 
\end{pmatrix}_{\bar{\mathbf{2}}},\qquad
(w)_{\mathbf{1}'}\otimes 
\begin{pmatrix}
x_1 \\
x_2 
\end{pmatrix}_{\bar{\mathbf{2}}}
=\begin{pmatrix}
wx_1 \\
-wx_2 
\end{pmatrix}_{\bar{\mathbf{2}}},
\end{aligned}
\end{equation}
where $\bar{\mathbf{2}}=\mathbf{2}, \mathbf{2}'$. The two doublets $x=(x_1,x_2)^\intercal$ and $y=(y_1,y_2)^\intercal$ yield
\begin{equation}
\label{D5C22}
\begin{aligned}
&\begin{pmatrix}
x_1 \\
x_2 
\end{pmatrix}_{\mathbf{2}}\otimes 
\begin{pmatrix}
y_1 \\
y_2 
\end{pmatrix}_{\mathbf{2}}
=(x_1y_2+x_2y_1)_{\mathbf{1}}
\oplus (x_1y_2-x_2y_1)_{\mathbf{1}'}
\oplus 
\begin{pmatrix}
x_1y_1 \\
x_2y_2 
\end{pmatrix}_{\mathbf{2}'}, \\&
\begin{pmatrix}
x_1 \\
x_2 
\end{pmatrix}_{\mathbf{2}'}\otimes 
\begin{pmatrix}
y_1 \\
y_2 
\end{pmatrix}_{\mathbf{2}'}
=(x_1y_2+x_2y_1)_{\mathbf{1}}
\oplus (x_1y_2-x_2y_1)_{\mathbf{1}'}
\oplus 
\begin{pmatrix}
x_2y_2 \\
x_1y_1 
\end{pmatrix}_{\mathbf{2}}, \\&
\begin{pmatrix}
x_1 \\
x_2 
\end{pmatrix}_{\mathbf{2}}\otimes 
\begin{pmatrix}
y_1 \\
y_2 
\end{pmatrix}_{\mathbf{2}'}
=\begin{pmatrix}
x_2y_1 \\
x_1y_2
\end{pmatrix}_{\mathbf{2}} 
\oplus 
\begin{pmatrix}
x_2y_2 \\
x_1y_1
\end{pmatrix}_{\mathbf{2}'}. 
\end{aligned}
\end{equation}

\subsection{Scalar potential}

In the 4HDM, there are four scalar fields $\phi_1$, $\phi_2$, $\phi_3$, and $\phi_4$, all of which are $SU(2)$ doublets. We introduce a $D_5$ symmetry into the model and construct the four-component vector $\vec{\Phi} = (\phi_1,\phi_2,\phi_3,\phi_4)^T$ in the internal space, comprising the Higgs-field components. We assume that $(\phi_1,\phi_2)$ transforms as a doublet under $D_5$ and $(\phi_3,\phi_4)$ forms another doublet. The transformations of the scalar fields $\phi_i$ under the $D_5$ symmetry are then given by:
\begin{equation}
\label{Phi1234}
\begin{pmatrix}
\phi _1    \\
\phi _2    \\
\end{pmatrix}\sim \mathbf{2},\quad \quad 
\begin{pmatrix}
\phi _3    \\
\phi _4    \\
\end{pmatrix}\sim \mathbf{2}'.
\end{equation}
The most general scalar potential of the $D_5$-symmetric 4HDM, consistent with the complex representation given in Eq.~(\ref{D5Cg2}), consists of quadratic and quartic terms:
\begin{equation}
\label{VC}
V_{D_5}=V_{2}+V_{4}.
\end{equation}
Using the tensor products given in Eqs.~(\ref{D511})--(\ref{D5C22}) for the complex representation, we compute $\Phi^\dagger \otimes \Phi$ and extract the combinations of $\phi^\dagger_i\phi_j$ that transform as the invariant singlet representation $\mathbf{1}$. Thus, the quadratic part of the $D_5$-invariant potential is given by
\begin{equation}
\label{VC2}
V_{2}=-\mu^2_{1}(\phi^\dagger_2\phi_2+\phi^\dagger_1\phi_1)-\mu^2_{2}(\phi^\dagger_4\phi_4+\phi^\dagger_3\phi_3).
\end{equation}
By evaluating the tensor product $({\Phi^\dagger}\otimes{\Phi})\otimes({\Phi^\dagger}\otimes{\Phi})$, we obtain the quartic part of the $D_5$-invariant potential:
\begin{equation}
\begin{aligned}
\label{Vl4}
V_{4} = & l_1(\phi^\dagger_2\phi_2 + \phi^\dagger_1 \phi_1)^2+ l_2(\phi^\dagger_2\phi_2 - \phi^\dagger_1\phi_1)^2
+ l_3(\phi^\dagger_4\phi_4 + \phi^\dagger_3\phi_3)^2
+l_4(\phi^\dagger_4\phi_4-\phi^\dagger_3\phi_3)^2\\
&+l_5(\phi^\dagger_2\phi_2+\phi^\dagger_1\phi_1)(\phi^\dagger_4\phi_4+\phi^\dagger_3\phi_3)
+l_6(\phi^\dagger_2\phi_2-\phi^\dagger_1\phi_1)(\phi^\dagger_4\phi_4-\phi^\dagger_3\phi_3)\\
&+l_7(\phi^\dagger_1\phi_2)(\phi^\dagger_2\phi_1)
+l_8(\phi^\dagger_3\phi_4)(\phi^\dagger_4\phi_3)+l_9\left[(\phi^\dagger_1\phi_3)(\phi^\dagger_3\phi_1)+(\phi^\dagger_2\phi_4)(\phi^\dagger_4\phi_2)\right]\\
&+l_{10}\left[(\phi^\dagger_1\phi_4)(\phi^\dagger_4\phi_1)+(\phi^\dagger_2\phi_3)(\phi^\dagger_3\phi_2)\right] +\left\{ 
l_{11}(\phi^\dagger_1\phi_3)(\phi^\dagger_2\phi_4)
+l_{12}(\phi^\dagger_2\phi_3)(\phi^\dagger_1\phi_4)
\right.\\
&\left.+l_{13}\left[(\phi^\dagger_2\phi_1)(\phi^\dagger_2\phi_3)+(\phi^\dagger_1\phi_2)(\phi^\dagger_1\phi_4)\right]
+l_{14}\left[(\phi^\dagger_1\phi_3)(\phi^\dagger_4\phi_3)+(\phi^\dagger_2\phi_4)(\phi^\dagger_3\phi_4)\right]
+\text{h.c.} \right\}.
\end{aligned}
\end{equation}
We now introduce an alternative notation for the scalar potential. Expanding the first six terms in Eq.~(\ref{Vl4}) and combining like terms yields new contributions whose coefficients are combinations of the original parameters. We define these combinations as new parameters: 
\begin{equation}
\begin{aligned}
&\lambda_1=l_1+l_2,\quad
\lambda_2=l_3+l_4,\quad
\lambda_3=2(l_1-l_2),\quad  
\lambda_4=2(l_3-l_4), \\&
\lambda_5=l_5+l_6,\quad
\lambda_6=l_5-l_6,\quad
\lambda_{\alpha}=l_{\alpha}, (\alpha=7,8,...,14).
\end{aligned}
\end{equation}
In terms of the new parameters, the invariant quartic part of the potential can be written as:
\begin{equation}
\begin{aligned}
\label{V4}
V_4=&\lambda_1\left[(\phi^\dagger_1\phi_1)^2+(\phi^\dagger_2\phi_2)^2\right]+
\lambda_2\left[(\phi^\dagger_3\phi_3)^2+(\phi^\dagger_4\phi_4)^2\right]+\lambda_3(\phi^\dagger_1\phi_1)(\phi^\dagger_2 \phi_2)+\lambda_4(\phi^\dagger_3\phi_3)(\phi^\dagger_4 \phi_4)\\
&+\lambda_5\left[(\phi^\dagger_1\phi_1)(\phi^\dagger_3\phi_3)+(\phi^\dagger_2\phi_2)(\phi^\dagger_4\phi_4)\right]+\lambda_6\left[(\phi^\dagger_2\phi_2)(\phi^\dagger_3\phi_3)+(\phi^\dagger_1\phi_1)(\phi^\dagger_4\phi_4)\right]\\
&+\lambda_{7}(\phi^\dagger_1\phi_2)(\phi^\dagger_2\phi_1)
+\lambda_{8}(\phi^\dagger_3\phi_4)(\phi^\dagger_4\phi_3)+\lambda_{9}\left[(\phi^\dagger_1\phi_3)(\phi^\dagger_3\phi_1)+(\phi^\dagger_2
\phi_4)(\phi^\dagger_4\phi_2)\right]\\
&+\lambda_{10}\left[(\phi^\dagger_1\phi_4)(\phi^\dagger_4\phi_1)+(\phi^\dagger_2\phi_3)(\phi^\dagger_3\phi_2)\right] +\left\{ 
\lambda_{11}(\phi^\dagger_1\phi_3)(\phi^\dagger_2\phi_4)
+\lambda_{12}(\phi^\dagger_2\phi_3)(\phi^\dagger_1\phi_4)
\right.\\
&\left.+\lambda_{13}\left[(\phi^\dagger_2\phi_1)(\phi^\dagger_2\phi_3)+(\phi^\dagger_1\phi_2)(\phi^\dagger_1\phi_4)\right]
+\lambda_{14}\left[(\phi^\dagger_1\phi_3)(\phi^\dagger_4\phi_3)+(\phi^\dagger_2\phi_4)(\phi^\dagger_3\phi_4)\right]
+\text{h.c.} \right\}.
\end{aligned}
\end{equation}
It is straightforward to verify that Eq.~(\ref{V4}) is equivalent to Eq.~(\ref{Vl4}). In this work, we adopt the notation in Eq.~(\ref{V4}). The dihedral group $D_5$ satisfies $D_5 \simeq Z_5 \rtimes Z_2$, and its generators are:
\begin{equation}
a = \left( \begin{array}{cccc}
e^{\frac{2i\pi}{5}} & 0 & 0 & 0\\
0 & e^{-\frac{2i\pi}{5}} & 0 & 0\\
0 & 0 & e^{\frac{4i\pi}{5}} & 0\\
0 & 0 & 0 & e^{-\frac{4i\pi}{5}}
\end{array} \right),\quad
b =\left( \begin{array}{cccc}
0 & 1 & 0 & 0\\
1 & 0 & 0 & 0\\
0 & 0 & 0 & 1\\
0 & 0 & 1 & 0
\end{array} \right).
\label{D5gab}
\end{equation}

It should be emphasized that if two field configurations $(\phi_1, \phi_2, \phi_3,\phi_4 )$ and $ (\phi_1', \phi_2', \phi_3',\phi_4')$ are related by a $Z_5$ or $Z_2$ transformation within $D_5$ (i.e., one can be obtained from the other by applying such a symmetry operation), then, owing to the invariance of the scalar potential under $D_5$ transformations acting on the Higgs fields, the two configurations are physically equivalent. Hence, it suffices to consider only one representative from each equivalence class in the vacuum analysis. 

We require all 16 unknown parameters in the potential to be real, implying that there is no explicit CP violation in the potential. This allows us to study the scalar sector under CP conservation and to investigate spontaneous CP violation induced by complex vevs. 

In Ref.~\cite{Hagedorn:2006ir}, an alternative notation was adopted to construct the general $D_5$-symmetric 4HDM scalar potential, and it was shown that this potential contains no accidental symmetries. We now demonstrate that the scalar potential in the notation used for the $D_5$ 4HDM in this work likewise contains no accidental symmetries. We assume that the fields $\phi_i$ ($i = 1, 2, 3, 4$) transform as follows:
\begin{equation}
    \phi_1\to\phi_1e^{i\sigma_1}, \quad\phi_2\to\phi_2e^{i\sigma_2}, \quad\phi_3\to\phi_3e^{i\sigma_3}, \quad\phi_4\to\phi_4e^{i\sigma_4}.
\end{equation}
Then the parameter terms $\mu^2_{1,2}$ and $\lambda_{k}$ $(k=1,2,...,10)$ all preserve the full $U(1)^4$ symmetry. When the condition $\sigma_1=\sigma_3+\sigma_4-\sigma_2$ is satisfied, each of the $\lambda_{11}$ and $\lambda_{12}$ terms individually breaks the $U(1)^4$ symmetry down to the same $U(1)^3$ symmetry. Meanwhile, $\lambda_{13}$ and $\lambda_{14}$ are constrained, respectively, by the conditions $2\sigma_2=\sigma_1+\sigma_3, 2\sigma_1=\sigma_2+\sigma_4$ and $2\sigma_3=\sigma_1+\sigma_4, 2\sigma_4=\sigma_2+\sigma_3$, which can further break $U(1)^3$ to $U(1)^2$. In other words, none of the potential parameter terms is invariant only under the $U(1)_Y$ gauge symmetry; each must also respect at least one additional $U(1)$ symmetry. Furthermore, only when $\lambda_{13} \neq 0$ and $\lambda_{14} \neq 0$, and under the condition $\sigma_1=\sigma_2=\sigma_3=\sigma_4$, can the $U(1)^4$ symmetry be broken to the $U(1)_Y$ gauge symmetry. In other words, even if the potential contains both the $\lambda_{11}$ and $\lambda_{12}$ terms together with only the $\lambda_{13}$ term (or only the $\lambda_{14}$ term), it still remains invariant under $U(1)^2$. Therefore, when $\lambda_{13}=0$ or $\lambda_{14}=0$, an accidental $U(1)$ symmetry inevitably appears in the scalar potential in addition to the gauge symmetry $U(1)_Y$. This leads to serious consequences: first, in the vacuum structure, the corresponding vevs are no longer minima of the scalar potential of the $D_5$ 4HDM; second, phenomenologically unacceptable massless Goldstone bosons emerge in the scalar mass spectrum. Consequently, in a 4HDM with an exact $D_5$ symmetry, it is necessary to satisfy:
\begin{equation}
    \lambda_{13}\neq0,\quad\text{and} \quad \lambda_{14}\neq0.
\end{equation}
This is the most fundamental prerequisite for the model's validity.
\subsection{Bounded from below limits}

To guarantee a stable minimum in the $D_5$ 4HDM, the scalar potential must be bounded from below (BFB); that is, there must be no direction in field space along which the potential tends to negative infinity. Since this paper focuses on the neutral vacuum structure, we consider only the bounded-from-below conditions along neutral directions (BFB-n).
To ensure the stability of the $D_5$ 4HDM potential, we examine all possible directions in which the fields $\phi_1$, $\phi_2$, $\phi_3$, and $\phi_4$ tend to infinity. The quadratic parameters $\mu^2_{1}$ and $\mu^2_{2}$ in the potential must be positive. To derive the necessary conditions for the quartic parameters $\lambda_i$, a straightforward approach is to study the behavior of the potential along specific field directions—namely, for any direction in field space along which $\phi_i \to \infty$, one requires that $V_4 > 0$\cite{Branco:2011iw}. Following the methods of Refs.~\cite{Ferreira:2004yd,Ferreira:2009jb,Grzadkowski:2009bt,Kannike:2012pe} for deriving the BFB conditions in the 2HDM and 3HDM, we derive a set of necessary and sufficient BFB-n conditions for the $D_5$ 4HDM:
\begin{equation}
\label{BFBn1}
\begin{aligned}
&  \lambda_{1}>0,\quad \lambda_{2}>0,\quad 
    \lambda_{w}\ge-2 \lambda_{1},\quad
    \lambda_{x}\ge-2 \lambda_{2},\\&  
    \lambda_{y}\ge-2 \sqrt{\lambda_{1}\lambda_{2}},\quad \lambda_{z}\ge-2 \sqrt{\lambda_{1}\lambda_{2}},
\end{aligned}
\end{equation}
where
\begin{equation}
\begin{aligned}
&\lambda_{w}=\lambda_{3}+\min(0,\lambda_{7}),\quad
\lambda_{x}=\lambda_{4}+\min(0,\lambda_{8}),\\ &
\lambda_{y}=\lambda_{5}+\min(0,\lambda_{9}),\quad
\lambda_{z}=\lambda_{6}+\min(0,\lambda_{10}),
\end{aligned}
\end{equation}
and the necessary conditions:
\begin{equation}
\label{BFBn2}
    \quad \left |\lambda_{11}+\lambda_{12} \right |+ 
2\left |\lambda_{13} +\lambda_{14} \right |<\lambda_a, 
\end{equation}
where
\begin{equation}
\lambda_a=\lambda_1+\lambda_2+\frac{\lambda_{3}}{2}+\frac{\lambda_{4}}{2}+\lambda_{5}+\lambda_{6}+\frac{\lambda_{7}}{2}+\frac{\lambda_{8}}{2}+\lambda_{9}+\lambda_{10}.
\end{equation}
These conditions ensure the stability of the scalar potential along the neutral directions.

\section{General properties of vacua}
\label{Sec:vacua}

After spontaneous symmetry breaking (SSB), we express the complex scalar field as an $SU(2)$ doublet as follows:
\begin{equation}
\phi_i=\left(\begin{array}{c}
h_i^{+} \\
\frac{1}{\sqrt{2}}\left(u_i+h_i+i a_i\right)
\end{array}\right), \quad i=1,2,3,4,
\end{equation}
where $u_i = v_i e^{i\theta_i}$, with real moduli $v_i$ and arbitrary phases $\theta_i$, and $\sqrt{v_1^2+v_2^2+v_3^2+v_4^2}=v_w=246$ GeV. When all phases $\theta_i$ vanish, the vacuum is real. Since a global $U(1)$ rotation can absorb one overall phase, we fix $\theta_4 = 0$ without loss of generality. If, after this rotation, any of the remaining phases $\theta_i$ is nonzero, the vacuum is complex.  

The purpose of this paper is to identify all possible complex vacuum states in the $D_5$ 4HDM that can give rise to SCPV. A vacuum state corresponds to a specific configuration of vacuum expectation values (vevs) for four complex Higgs doublets $\phi_i$ $(i = 1, 2, 3, 4)$, which can be written as $(v_1e^{i\theta_1}, v_2e^{i\theta_2}, v_3e^{i\theta_3}, v_4)/\sqrt{2}$. We determine these states by minimizing the scalar potential with respect to $v_i$ and $\theta_i$. For notational simplicity, we omit the $\sqrt{2}$ factor in the text but include it in all calculations. 

We further classify all possible vacuum solutions obtained from the minimization of the potential according to the following definitions:
\begin{enumerate}
    \item \textbf{Normal vacua:} The vacuum expectation values (vevs) of all four doublets are nonzero ($v_i \neq 0$), so each Higgs doublet participates in electroweak symmetry breaking.

   \item \textbf{Inert vacua:} At least one doublet has a vanishing vacuum expectation value (VEV), $v_i = 0$. The corresponding doublet does not participate in electroweak symmetry breaking, thereby defining an inert model.
\end{enumerate}
Notice that, unlike transforming to the Higgs basis, inert vacua are identified in the original $D_5$-symmetric basis, in which the symmetry remains manifest. In the Higgs basis, by contrast, the four doublets are recombined so that all nonzero vevs are rotated into a single doublet, artificially yielding a representation in which the other three vevs vanish. This basis change merely alters the field coordinates, rendering the original symmetry non-manifest; it does not correspond to a different physical vacuum. For a detailed definition and discussion, see Ref.~\cite{Branco:2011iw}.

\subsection{Stationarity conditions}

Our first step is to determine all stationary points of the scalar potential associated with the neutral vacuum, i.e., the vevs that satisfy the first-derivative conditions of the potential. We consider the potential $V(u_i, u_i^\ast)$ and solve for the vevs by imposing the stationarity conditions:
\begin{equation}
\frac{\partial V}{\partial u_i}=0,\quad \frac{\partial V}{\partial u_i^\ast}=0, \quad i=1,2,3,4.
\end{equation}
These two conditions are complex conjugates of one another and therefore equivalent; hence, it suffices to use only one of them. Choosing the latter yields the following four stationarity conditions:
\begin{equation}
\begin{aligned}
   \frac{\partial V}{\partial u^\ast_1}=&-\frac{1}{2}\mu^2_1u_1+\frac{1}{2}\lambda_1u_1|u_1|^2+\frac{1}{4}u_1 \left[(\lambda_3+\lambda_7)|u_2|^2+(\lambda_5+\lambda_9)|u_3|^2+(\lambda_6+\lambda_{10})|u_4|^2\right]\\
   &+\frac{1}{4}\left[(\lambda_{11}+\lambda_{12})u^\ast_2u_3u_4+\lambda_{13}(2u^\ast_1u_2u_4+u^\ast_3u^2_2)+\lambda_{14}u^\ast_4u^2_3\right]=0,
   \label{Vu1}
\end{aligned}
\end{equation}
\begin{equation}
\begin{aligned}
   \frac{\partial V}{\partial u^\ast_2}=&-\frac{1}{2}\mu^2_1u_2+\frac{1}{2}\lambda_1u_2|u_2|^2+\frac{1}{4}u_2\left[(\lambda_3+\lambda_7)|u_1|^2+(\lambda_5+\lambda_9)|u_4|^2+(\lambda_6+\lambda_{10})|u_3|^2\right]\\
   &+\frac{1}{4}\left[(\lambda_{11}+\lambda_{12})u^\ast_1u_3u_4+\lambda_{13}(2u^\ast_2u_1u_3+u^\ast_4u^2_1)+\lambda_{14}u^\ast_3u^2_4\right]=0,
    \label{Vu2}
\end{aligned}
\end{equation}
\begin{equation}
\begin{aligned}
   \frac{\partial V}{\partial u^\ast_3}=&-\frac{1}{2}\mu^2_2u_3+\frac{1}{2}\lambda_2u_3|u_3|^2+\frac{1}{4}u_3\left[(\lambda_4+\lambda_8)|u_4|^2+(\lambda_5+\lambda_9)|u_1|^2+(\lambda_6+\lambda_{10})|u_2|^2\right]\\
   &+\frac{1}{4}\left[(\lambda_{11}+\lambda_{12})u^\ast_4u_1u_2+\lambda_{13}u^\ast_1u^2_2+\lambda_{14}(2u^\ast_3u_1u_4+u^\ast_2u^2_4)\right]=0,
   \label{Vu3}
\end{aligned}
\end{equation}
\begin{equation}
\begin{aligned}
   \frac{\partial V}{\partial u^\ast_4}=&-\frac{1}{2}\mu^2_2u_4+\frac{1}{2}\lambda_2u_4|u_4|^2+\frac{1}{4}u_4\left[(\lambda_4+\lambda_8)|u_3|^2+(\lambda_5+\lambda_9)|u_2|^2+(\lambda_6+\lambda_{10})|u_1|^2\right]\\
   &+\frac{1}{4}\left[(\lambda_{11}+\lambda_{12})u^\ast_3u_1u_2+\lambda_{13}u^\ast_2u^2_1+\lambda_{14}(2u^\ast_4u_2u_3+u^\ast_1u^2_3)\right]=0.
    \label{Vu4}
\end{aligned}
\end{equation}
Since the five parameter pairs ($\lambda_3, \lambda_7$), ($\lambda_4, \lambda_8$), ($\lambda_5, \lambda_9$), ($\lambda_6, \lambda_{10}$), and ($\lambda_{11}, \lambda_{12}$) enter the stationarity conditions only through their sums, there are effectively 11 independent parameters out of the original 16. To make the equations more concise, we introduce the following abbreviations:
\begin{equation}
\begin{aligned}
&\bar\lambda_{3}=\lambda_3+\lambda_7,\quad \bar\lambda_{4}=\lambda_4+\lambda_8,\quad
     \bar\lambda_{5}=\lambda_5+\lambda_9,\\&
      \bar\lambda_{6}=\lambda_6+\lambda_{10},\quad
       \bar\lambda_{11}=\lambda_{11}+\lambda_{12}.
\end{aligned}
\end{equation}
Although the stationarity conditions above are presented in complex form, they can equivalently be expressed in terms of the real parameters $v_i$ and $\theta_i$:
\begin{equation}
\frac{\partial V}{\partial v_i} = 0, \quad \frac{\partial V}{\partial \theta_i} = 0,\quad i=1,2,3,4.
\label{vtheta}
\end{equation}
After fixing $\theta_4=0$, the seven stationarity conditions are given in Appendix~B.

\subsection{Positive definiteness of the Hessian}

Subsequently, to ensure that the vevs obtained by solving the stationarity conditions correspond to local minima of the scalar potential, we construct the Hessian with respect to the VEV parameters:
\begin{equation}
H_C =\left(\begin{array}{cc}
\frac{\partial^2 V}{\partial v_i \partial v_j} & \frac{\partial^2 V}{\partial v_i \partial \theta_j} \\
\frac{\partial^2 V}{\partial \theta_i \partial v_j} & \frac{\partial^2 V}{\partial \theta_i \partial \theta_j}
\end{array}\right),
\label{Hessianc}
\end{equation}
and we require it to be positive definite. For real vacua, the Hessian is given by:
\begin{equation}
H_R =\left(\begin{array}{c}
\frac{\partial^2 V}{\partial v_i \partial v_j}
\end{array}\right).
\label{Hessianr}
\end{equation}

Finally, after obtaining all possible vevs that satisfy the model conditions, we determine whether each complex vacuum state leads to spontaneous CP violation.

\section{Real vacua}
\label{Sec:Realvacua}
\subsection{Real vevs}
\label{Sec:Real}
When all $\theta_i = 0$, the vacuum is real, and the vevs can be denoted as:
\begin{equation}
\label{rv1234}
(u_1,u_2,u_3,u_4)= (v_1, v_2, v_3, v_4).
\end{equation}

$\bullet$ \textbf{Normal vacua}

We now discuss the Normal vacua in the real case, in which all $v_i$ are nonzero. Substituting Eq.~(\ref{rv1234}) into the stationarity conditions (\ref{Vu1})--(\ref{Vu4}), we obtain the following four expressions for $\mu^2_1$ and $\mu^2_2$ in terms of the quartic couplings:
\begin{equation}
\label{Rmu1v1}
  \mu^2_1=\frac{1}{2}\left[2\lambda_1v_1^2+\bar\lambda_{3}v_2^2+\bar\lambda_{5}v_3^2+\bar\lambda_{6}v_4^2+\bar\lambda_{11}\frac{v_2v_3v_4}{v_1}+\lambda_{13}\left(2v_2v_4+\frac{v_2^2v_3}{v_1}\right)+\lambda_{14}\frac{v_3^2v_4}{v_1}\right],
\end{equation}
\begin{equation}
\label{Rmu1v2}
   \mu^2_1=\frac{1}{2}\left[2\lambda_1v_2^2+\bar\lambda_{3}v_1^2+\bar\lambda_{5}v_4^2+\bar\lambda_{6}v_3^2+\bar\lambda_{11}\frac{v_1v_3v_4}{v_2}+\lambda_{13}\left(2v_1v_3+\frac{v_1^2v_4}{v_2}\right)+\lambda_{14}\frac{v_3v_4^2}{v_2}\right],
\end{equation}
\begin{equation}
\label{Rmu2v3}
    \mu^2_2=\frac{1}{2}\left[2\lambda_2v_3^2+\bar\lambda_{4}v_4^2+\bar\lambda_{5}v_1^2+\bar\lambda_{6}v_2^2+\bar\lambda_{11}\frac{v_1v_2v_4}{v_3}+\lambda_{13}\frac{v_1v_2^2}{v_3}+\lambda_{14}\left(2v_1v_4+\frac{v_2v_4^2}{v_3}\right)\right],
\end{equation}
\begin{equation}
\label{Rmu2v4}
    \mu^2_2=\frac{1}{2}\left[2\lambda_2v_4^2+\bar\lambda_{4}v_3^2+\bar\lambda_{5}v_2^2+\bar\lambda_{6}v_1^2+\bar\lambda_{11}\frac{v_1v_2v_3}{v_4}+\lambda_{13}\frac{v_1^2v_2}{v_4}+\lambda_{14}\left(2v_2v_3+\frac{v_1v_3^2}{v_4}\right)\right].
\end{equation}
Note that equations (\ref{Rmu1v1}) and (\ref{Rmu1v2}) do not hold when $v_1 = 0$ and $v_2 = 0$, respectively. Similarly, equations (\ref{Rmu2v3}) and (\ref{Rmu2v4}) do not hold when $v_3 = 0$ and $v_4 = 0$, respectively. Since the two expressions for $\mu_1^2$ in (\ref{Rmu1v1}) and (\ref{Rmu1v2}) must be mutually consistent, and likewise for $\mu_2^2$ in (\ref{Rmu2v3}) and (\ref{Rmu2v4}), we obtain the following two consistency conditions:
\begin{subequations}
\begin{equation}
\begin{aligned}
\label{rmu1mmu1}
&\left(2\lambda_1-\bar\lambda_{3}-\bar\lambda_{11}\frac{v_3v_4}{v_1v_2}\right)(v^2_1-v^2_2)+\left(\bar\lambda_{5}-\bar\lambda_{6}\right)(v^2_3-v^2_4)\\
&+\lambda_{13}\left(\frac{v_3(v_2^2-2v_1^2)}{v_1}+\frac{v_4(2v^2_2-v_1^2)}{v_2}\right)+\lambda_{14}v_3v_4\left(\frac{v_3}{v_1}-\frac{v_4}{v_2}\right)=0,
\end{aligned}
\end{equation}
\begin{equation}
\begin{aligned}
\label{rmu2mmu2}
&\left(2\lambda_2-\bar\lambda_{4}-\bar\lambda_{11}\frac{v_1v_2}{v_3v_4}\right)(v^2_3-v^2_4)+(\bar\lambda_{5}-\bar\lambda_{6})(v^2_1-v^2_2)\\
&+\lambda_{13}v_1v_2\left(\frac{v_2}{v_3}-\frac{v_1}{v_4}\right)+\lambda_{14}\left(\frac{v_2(v^2_4-2v_3^2)}{v_3}+\frac{v_1(2v_4^2-v_3^2)}{v_4}\right)=0.
\end{aligned}
\end{equation}
\end{subequations}
First, consider the most general real vacuum ($v_1, v_2, v_3, v_4$). We find that, provided the BFB conditions (\ref{BFBn1}) and (\ref{BFBn2}) hold, it is possible to satisfy the two consistency conditions (\ref{rmu1mmu1}) and (\ref{rmu2mmu2}) by appropriately choosing the remaining potential parameters $\lambda_1$, $\lambda_2$, $\bar\lambda_3$, $\bar\lambda_4$, $\bar\lambda_5$, $\bar\lambda_6$, and $\bar\lambda_{11}$ so that $ \lambda_{13} \neq 0 $ and $\lambda_{14} \neq 0 $. Therefore, the vacuum $(v_1, v_2, v_3, v_4)$ constitutes a solution of the model, subject to the constraints on the potential parameters imposed by Eqs.~(\ref{Rmu1v1}), (\ref{Rmu2v3}), (\ref{rmu1mmu1}), and (\ref{rmu2mmu2}). 

Next, we consider special cases in which some of the moduli $|v_i|$ of the vacuum $(v_1, v_2, v_3, v_4)$ are equal, to determine whether these equalities can simplify the constraints on the potential parameters or reduce their number. The following cases are considered:

\begin{itemize}
\item[(1)] 
When $\left |v_1 \right |=\left |v_2 \right |$ and $\left |v_3 \right |=\left |v_4 \right |$, the vevs can take four forms: $(v_1, v_1, v_3, v_3)$, $(v_1, -v_1, -v_3, v_3)$, $(v_1, -v_1, v_3, v_3)$, and $(v_1, v_1, -v_3, v_3)$. 
However, the last two imply $\lambda_{13}=0$ and $\lambda_{14}=0$, which do not satisfy the model requirements. Thus, only the first two forms are acceptable. For simplicity, we denote them uniformly as the vacuum $(v_1,\pm v_1,\pm v_3,v_3)$, since both sides of the constraint Eqs.~(\ref{rmu1mmu1}) and (\ref{rmu2mmu2}) vanish identically; consequently, the number of effective constraints on the parameters is reduced from four to two.

\item[(2)] 
When $\left |v_1 \right |=\left |v_3 \right |$ and $\left |v_2 \right |=\left |v_4 \right |$, the vevs can be expressed in four forms: $(v_1, v_2, v_1, v_2)$, $(v_1, -v_2, -v_1, v_2)$, $(v_1, -v_2, v_1, v_2)$, and $(v_1, v_2, -v_1, v_2)$. None of these forms imply $\lambda_{13}=0$ or $\lambda_{14}=0$, and, while they significantly simplify the constraints, their number remains four. We choose the first two forms for discussion, which we denote collectively as the vacuum $(v_1, \pm v_2, \pm v_1, v_2)$.

\item[(3)] 
When $\left |v_1 \right |=\left |v_4 \right |$ and $\left |v_2 \right |=\left |v_3 \right |$, this case is similar to the one above. We select two forms for discussion, denoted collectively as the vacuum $(v_1, \pm v_2, v_2, \pm v_1)$.

\item[(4)] 
When any three moduli $|v_i|$ are equal, there are four cases up to permutation. Taking $|v_1| = |v_2| = |v_3| = v$ as an example, the vevs can be expressed in four forms: $(v,v,v,v_4)$, $(v,-v,-v,v_4)$, $(v,-v,v,v_4)$, and $(v,v,-v,v_4)$. None of these forms imply $\lambda_{13}=0$ or $\lambda_{14}=0$. Since there are numerous permutations of such special vacua, and since the constraints on the potential parameters are neither significantly simplified nor reduced in number, and the same holds for the complex vacua discussed later, we do not further elaborate on these special vacua in this paper.

\item[(5)] 
When $|v_1| = |v_2| = |v_3| = |v_4| = v$ and $\lambda_{13} \neq 0$, $\lambda_{14} \neq 0$, the vevs reduce to two forms, $(v,v,v,v)$ and $(v,-v,-v,v)$, which we denote uniformly as $(v,\pm v,\pm v,v)$. In this case, the two constraints on the potential parameters can be obtained directly from those for the vacuum $(v_1,\pm v_1,\pm v_3,v_3)$ by setting $v_1 = v_3 = v$, but their number is not further reduced. The same holds for the complex vacua discussed later, so we do not further elaborate on this special vacuum in this paper.
\end{itemize}
The constraints on the potential parameters for the first three special vacua discussed above are summarized in Table~\ref{Table:rvev}. \\

$\bullet$ \textbf{Inert vacua}

We now discuss the inert vacua among the real vacua. First, we consider the case in which only one of the $v_i$ is zero. Taking $v_1 = 0$ as an example, the vevs are given by $(0, v_2, v_3, v_4)$. In this case, Eq.~(\ref{Rmu1v1}) no longer applies; instead, we use the original stationarity condition $\frac{\partial V}{\partial v_1} = 0$ to obtain:
\begin{equation}
\label{rvv1}
 \lambda_{13}=-\frac{(\bar\lambda_{11}v_2+\lambda_{14}v_3)v_4}{v^2_2}.
\end{equation}
Furthermore, the expression for $\mu_1^2$ (\ref{Rmu1v2}) is automatically consistent, while the two expressions for $\mu_2^2$ (\ref{Rmu2v3}) and (\ref{Rmu2v4}) still need to satisfy the consistency condition (\ref{rmu2mmu2}). We find that this vacuum does not lead to $\lambda_{13}=0$ or $\lambda_{14}=0$ under these constraints and is a solution of the model. For other real vacua in which only one of the $v_i$ is zero, the situation is similar by permutation of the indices, and all such vacua satisfy the requirements of the model. Moreover, we have examined special cases where some of the $|v_i|$ are equal. We find that although the constraints can be simplified, their number is not reduced, and the same holds for the complex vacua discussed later; hence, we do not elaborate further on these special vacua in this paper.

Secondly, we consider the case where two of the $v_i$ are zero. Taking $v_1 = v_2 = 0$ as an example, the vevs can be denoted as $(0, 0, v_3, v_4)$. In this case, Eqs.~(\ref{Rmu1v1}) and (\ref{Rmu1v2}) no longer hold; instead, the original stationarity conditions $\frac{\partial V}{\partial v_1} = 0$ and $\frac{\partial V}{\partial v_2} = 0$ are used to obtain:
\begin{equation}
  \lambda_{14}v_4v_3^2=0, \quad\text{and} \quad \lambda_{14}v_3v_4^2=0.
\end{equation}
Further solving yields $\lambda_{14}=0$, so this vacuum fails to satisfy the model requirements. By explicit calculation, we find that for all real vacua in which two of the $v_i$ are zero, the stationarity conditions $\frac{\partial V}{\partial v_i}=0$ inevitably imply that at least one of $\lambda_{13}=0$ or $\lambda_{14}=0$. Therefore, in this case, no vacuum satisfies the model requirements.

Finally, we consider the case in which three of the $v_i$ are zero. Taking $v_2 = v_3 = v_4 = 0$ as an example, the vevs are $(v, 0, 0, 0)$. In this case, Eq.~(\ref{Rmu1v1}) yields the following unique constraint:
\begin{equation}
    \mu^2_1=\lambda_1v^2.
\end{equation}
It does not imply $\lambda_{13}=0$ or $\lambda_{14}=0$, so this vacuum is a valid solution of the model. For other real vacua in which three of the $v_i$ vanish, the situation is similar, and all such vacua satisfy the requirements of the model.

In Table~\ref{Table:rvev}, we summarize the complete list of all possible real vacua and the constraints on the parameter space that admit each solution. For convenience in comparing real and complex vacua, we classify each real vacuum in the table using the notation R‑X‑Yz. Here, R denotes a real vacuum, X is either N (normal vacuum) or I (inert vacuum), Y gives the number of constraints on the potential parameters, and z distinguishes cases with identical values of X and Y.
\begin{table}[htbp]
\centering
\begin{tabular}{|c|c|c|}
\hline 
\makecell[c]{Vacuum} &Real vevs& Constraints \\
\hline 
R-N-2 &$v_1,\pm v_1,\pm v_3, v_3$&\makecell[c]{$\mu^2_1=\frac{1}{2}\left[(2\lambda_1+\bar\lambda_{3})v_1^2+(\bar\lambda_{5}+\bar\lambda_{6}+\bar\lambda_{11})v_3^2\pm3\lambda_{13}v_1v_3+\lambda_{14}\frac{v_3^3}{v_1}\right]$,\\$\mu^2_2=\frac{1}{2}\left[(2\lambda_2+\bar\lambda_{4})v_3^2+(\bar\lambda_{5}+\bar\lambda_{6}+\bar\lambda_{11})v_1^2\pm\lambda_{13}\frac{v_1^3}{v_3}+3\lambda_{14}v_1v_3\right]$ }\\
\hline 
R-N-4a  &$v_1, v_2,v_3, v_4$&Eqs.~(\ref{Rmu1v1}), (\ref{Rmu2v3}), (\ref{rmu1mmu1}), (\ref{rmu2mmu2})\\
\hline 
R-N-4b  &$v_1,\pm v_2,\pm v_1, v_2$&\makecell[c]{$\mu^2_1=\frac{1}{2}[(2\lambda_1+\bar\lambda_{5})(v_1^2+v_2^2)+\lambda_{14}v_1v_2],$\\$\mu^2_2=\frac{1}{2}[(2\lambda_2+\bar\lambda_{5})(v_1^2+v_2^2)+\lambda_{14}v_1v_2]$,\\$\lambda_1=\frac{1}{2}(\bar\lambda_{3}-\bar\lambda_{5}+\bar\lambda_{6}+\bar\lambda_{11}\pm 3\lambda_{13})$,\\$\lambda_2=\frac{1}{2}\left(\bar\lambda_{4}-\bar\lambda_{5}+\bar\lambda_{6}+\bar\lambda_{11}\pm\lambda_{13}+\lambda_{14}\frac{v_1^2+v_2^2}{v_1v_2}\right)$}\\
\hline 
R-N-4c  &$v_1,\pm v_2, v_2, \pm v_1$&\makecell[c]{$\mu^2_1=\frac{1}{2}[(2\lambda_1+\bar\lambda_{6})(v_1^2+v_2^2)+\lambda_{13}v_1v_2],$\\$\mu^2_2=\frac{1}{2}[(2\lambda_2+\bar\lambda_{6})(v_1^2+v_2^2)+\lambda_{13}v_1v_2]$,\\$\lambda_1=\frac{1}{2}\left(\bar\lambda_{3}+\bar\lambda_{5}-\bar\lambda_{6}+\bar\lambda_{11}\pm\lambda_{14}+\lambda_{13}\frac{v_1^2+v_2^2}{v_1v_2}\right)$,\\$\lambda_2=\frac{1}{2}(\bar\lambda_{4}+\bar\lambda_{5}-\bar\lambda_{6}+\bar\lambda_{11}\pm3\lambda_{14})$}\\
\hline 
R-I-1a  &$v, 0, 0,0$ &\makecell[c]{ $\mu_1^2=\lambda_1 v^2$} \\
\hline 
R-I-1b  &$0, 0, 0,v$ &\makecell[c]{ $\mu_2^2=\lambda_2 v^2$} \\
\hline 
R-I-4a &$v_1, v_2, v_3,0$&\makecell[c]{$ 
\mu^2_1=\lambda_1v_2^2+\frac{1}{2}(\bar\lambda_{3}v_1^2+\bar\lambda_{6}v^2_3)+\lambda_{13}v_1v_3,$\\$\mu^2_2=\lambda_2v_3^2+\frac{1}{2}\left(\bar\lambda_{5}v_1^2+\bar\lambda_{6}v^2_2+\lambda_{13}\frac{v_1v_2^2}{v_3}\right),$\\$ \lambda_{13}\frac{v_3(2v_1^2-v_2^2)}{v_1}=(2\lambda_1-\bar\lambda_{3})(v_1^2-v_2^2)+(\bar\lambda_{5}-\bar\lambda_{6})v_3^2,$\\$\lambda_{14}=-\frac{(\bar\lambda_{11}v_3+\lambda_{13}v_1)v_2}{v_3^2}$} \\
\hline 
R-I-4b &$v_1, v_2, 0,v_4$&\makecell[c]{$ 
\mu^2_1=\lambda_1v_1^2+\frac{1}{2}(\bar\lambda_{3}v_2^2+\bar\lambda_{6}v^2_4)+\lambda_{13}v_2v_4,$\\$\mu^2_2=\lambda_2v_4^2+\frac{1}{2}\left(\bar\lambda_{5}v_2^2+\bar\lambda_{6}v^2_1+\lambda_{13}\frac{v_1^2v_2}{v_4}\right),$\\$ \lambda_{13}\frac{v_4(2v_2^2-v_1^2)}{v_2}=(2\lambda_1-\bar\lambda_{3})(v_2^2-v_1^2)+(\bar\lambda_{5}-\bar\lambda_{6})v_4^2,$\\$\lambda_{14}=-\frac{(\bar\lambda_{11}v_4+\lambda_{13}v_2)v_1}{v_4^2}$} \\
\hline 
R-I-4c &$v_1,0, v_3, v_4$&\makecell[c]{ $\mu^2_1=\lambda_1v_1^2+\frac{1}{2}\left(\bar\lambda_{5}v_3^2+\bar\lambda_{6}v_4^2+\lambda_{14}\frac{v_3^2v_4}{v_1}\right),$\\$ 
\mu^2_2=\lambda_2v_3^2+\frac{1}{2}(\bar\lambda_{4}v_4^2+\bar\lambda_{5}v^2_1)+\lambda_{14}v_1v_4,$\\$ \lambda_{13}=-\frac{(\bar\lambda_{11}v_1+\lambda_{14}v_4)v_3}{v_1^2} ,$\\$\lambda_{14}\frac{v_1(2v_4^2-v_3^2)}{v_4}=(2\lambda_2-\bar\lambda_{4})(v_4^2-v_3^2)+(\bar\lambda_{6}-\bar\lambda_{5})v_1^2$} \\
\hline 
R-I-4d &$0,v_2, v_3, v_4$&\makecell[c]{$\mu^2_1=\lambda_1v_2^2+\frac{1}{2}\left(\bar\lambda_{5}v_4^2+\bar\lambda_{6}v^2_3+\lambda_{14}\frac{v_3v_4^2}{v_2}\right) $,\\$\mu^2_2=\lambda_2v_4^2+\frac{1}{2}(\bar\lambda_{4}v_3^2+\bar\lambda_{5}v_2^2)+\lambda_{14}v_2v_3,$\\$\lambda_{13}=-\frac{(\bar\lambda_{11}v_2+\lambda_{14}v_3)v_4}{v^2_2}$,\\$\lambda_{14}\frac{v_2(2v_3^2-v_4^2)}{v_3}=(2\lambda_2-\bar\lambda_{4})(v_3^2-v_4^2)+(\bar\lambda_{6}-\bar\lambda_{5})v_2^2$} \\
\hline 
\end{tabular}
\caption{Real vacua. In the notation R-X-Yz, R indicates that the vacuum is real; X is either N (normal) or I (inert); and Y is the number of constraints on the potential parameters obtained by solving the stationarity conditions. The letter z distinguishes different vevs with the same X and Y values. Vacua labeled with $\pm$ come in two forms: the ``$+$'' case corresponds to one, and the ``$-$'' case to the other.
\label{Table:rvev}}
\end{table}

We now discuss each of the real vacua listed in Table \ref{Table:rvev}.
\begin{itemize}
	\item
Vacuum R-N-2 contains two forms: $(v_1,v_1,v_3,v_3)$ and $(v_1,-v_1,-v_3,v_3)$, with $v_1^2 + v_3^2 = \frac{v^2_{w}}{2}$. After SSB, the former preserves a residual $Z_2$ symmetry of the potential, whereas the latter completely breaks the $D_5$ symmetry.
\item
Vacuum R-N-4a is the most general real vacuum and completely breaks the $D_5$ symmetry.
\item
The Vacua R-N-4b and R-N-4c are related by a permutation of the Higgs fields, satisfy $v_1^2 + v_2^2 = \frac{v^2_{w}}{2}$, and completely break the $D_5$ symmetry.
\item
The Vacua R-I-1a and R-I-1b are also related by a permutation of the Higgs fields, satisfy $v = v_{w}$, and completely break the $D_5$ symmetry. In these two vacua, because only one Higgs doublet acquires a non-zero vev, there are no flavor-changing neutral currents (FCNCs).
\item
The Vacua R-I-4a, R-I-4b, R-I-4c, and R-I-4d satisfy, respectively, $\sqrt{v_1^2+v_2^2+v_3^2}=v_{w}$, $\sqrt{v_1^2+v_2^2+v_4^2} = v_w$, $\sqrt{v_1^2 + v_3^2 + v_4^2} = v_w$, and $\sqrt{v^2_2 + v^2_3 + v^2_4} = v_{w}$. All four vacua completely break the $D_5$ symmetry.
\end{itemize}

\paragraph{Relationships among real vacua}

To provide a more intuitive analysis of the real vacua listed in Table \ref{Table:rvev}, we further classify them into the following three types:

\begin{itemize}
\item[(1)] The first type comprises the Normal vacua R-N-2, R-N-4a, R-N-4b, and R-N-4c. Vacuum R-N-4a is the most general among them, as the other three can be obtained by assigning specific values to the parameters $v_i$ in R-N-4a.

\item[(2)] The second type comprises the Inert vacua R-I-1a and R-I-1b, with three of the $v_i$ set to zero. 

\item[(3)] The third type comprises the Inert vacua R-I-4a, R-I-4b, R-I-4c, and R-I-4d, with only one of the $v_i$ set to zero. 
\end{itemize}

\subsection{Positive definiteness of the Hessian}

We now discuss the conditions for the positive definiteness of the Hessians (\ref{Hessianr}) associated with these real vacua.

For the vacuum R-N-2 to be a local minimum of the potential, its Hessian must be positive definite, which requires that all four leading principal minors be positive. Accordingly, the potential parameters must satisfy the following four conditions:
\begin{equation}
\begin{aligned}
&\bullet \quad D_1=a_{11}>0,\\
&\bullet \quad D_2=a_{11}^2  - a_{12}^2>0 ,\\
&\bullet \quad  D_3=a_{33}\left(a_{11}^2- a_{12}^2\right)  - a_{11}\left(a_{13}^2 +a_{14}^2\right) + 2a_{12} a_{13}a_{14} > 0,\\
&\bullet \quad  D_4=\left(a_{11}^2 - a_{12}^2\right)\left(a_{33}^2 - a_{34}^2\right) - 2\left( a_{11} a_{33}+ a_{12} a_{34}\right) \left(a_{13}^2 + a_{14}^2\right) \\ &\quad \quad + 4 \left( a_{11}a_{34}+a_{12}  a_{33} \right)a_{13} a_{14} + \left(a_{13}^2 - a_{14}^2\right)^2>0,
  \label{RN2}
\end{aligned}
\end{equation}
where 
\begin{equation}
  \label{HRN2}
\begin{aligned}
& a_{11}=-\mu_1^2+3 \lambda_1 v_1^2+\frac{1}{2}\left[\bar{\lambda}_3 v_1^2+\left(\bar{\lambda}_5+\bar{\lambda}_6\right) v_3^2 \pm 2 \lambda_{13} v_1v_3\right], \\
& a_{33}=-\mu_2^2+3 \lambda_2 v_3^2+\frac{1}{2}\left[\bar{\lambda}_4 v_3^2+\left(\bar{\lambda}_5+\bar{\lambda}_6\right) v_1^2+2 \lambda_{14} v_1 v_3\right], \\
& a_{12}= \pm \bar{\lambda}_3 v_1^2 \pm \frac{\bar{\lambda}_{11}}{2}  v_3^2+2 \lambda_{13} v_1 v_3,\quad  a_{13}= \pm \left(\bar{\lambda}_5 + \frac{\bar{\lambda}_{11}}{2} \right) v_1 v_3+\frac{\lambda_{13}}{2}  v_1^2 \pm \lambda_{14} v_3^2,\\ & a_{14}=\left(\bar{\lambda}_6 +\frac{\bar{\lambda}_{11}}{2} \right) v_1 v_3 \pm \lambda_{13} v_1^2+\frac{\lambda_{14}}{2}  v_3^2,\quad  a_{34}= \pm \bar{\lambda}_4 v_3^2 \pm \frac{\bar{\lambda}_{11}}{2}  v_1^2\pm2 \lambda_{14}v_1v_3.
\end{aligned}
\end{equation}

For Vacuum R-I-1a, positive definiteness requires that:
\begin{equation}
 \lambda_1>0,\quad \bar\lambda_{3}>2\lambda_{1},\quad \bar\lambda_{5}>\frac{2\mu_2^2}{v^2},\quad \bar\lambda_{6}>\frac{2\mu_2^2}{v^2}.
\end{equation}

The positive-definiteness conditions for the Hessians that ensure the remaining vacua in Table \ref{Table:rvev} are local minima of the scalar potential are presented in Appendix~C.

\section{Complex vacua}
\label{Sec:Complexvacua}
\subsection{Spontaneous CP violation}

Before studying complex vacua, we first discuss how to determine whether a given vacuum leads to CP violation. By definition, in spontaneous CP violation induced by the vacuum, the Lagrangian before spontaneous symmetry breaking (SSB) is explicitly CP-conserving, which requires all parameters in the scalar potential to be real. In that case, if the vacuum after SSB is not CP-invariant, this indicates spontaneous CP violation. More precisely, if the Lagrangian is invariant under a CP transformation, yet there exists no transformation that can be physically identified with CP and that leaves both the vacuum and the Lagrangian invariant, then CP is spontaneously violated. In the SM, there is only one Higgs doublet, and its CP transformation is conventionally defined as $\phi(t, \vec{x}) \to \phi^\ast(t, -\vec{x})$, which requires all parameters to be real for the potential to be CP-invariant. Hence, no spontaneous CP violation arises from the scalar sector. In an extension of the SM, T. D. Lee~\cite{Lee:1973iz} constructed the two-Higgs-doublet model (2HDM) and proposed an early mechanism for spontaneous CP violation. We consider an MHDM consisting of $n$ $ SU(2)_L \times U(1)_Y$ Higgs doublets. One must consider the most general CP transformation that leaves the kinetic terms of the Lagrangian invariant, in order to accommodate all possible symmetries of the Lagrangian under which the Higgs doublets transform nontrivially. Hence, the CP transformation for $n$ Higgs doublets can be written as
\begin{equation}
	\phi_i \xrightarrow{\text{CP}} CP\phi_i (CP)^\dagger=\displaystyle\sum_{j=1}^{n} U_{ij} \phi_j^\ast ,
\label{cpt}
\end{equation}
where $U_{ij}$ denotes an arbitrary $n \times n $ unitary matrix acting on the Higgs doublets. Assuming that the vacuum is invariant under this CP transformation, i.e.,
\begin{equation}
CP |0\rangle = |0\rangle.
\label{cp0}
\end{equation}
By combining Eqs.~(\ref{cpt}) and (\ref{cp0}), the following relation can be obtained~\cite{Branco:1983tn}:
\begin{equation}
	\displaystyle\sum_{j=1}^{n} U_{ij} \langle 0|\phi_j|0\rangle^\ast=\langle 0|\phi_i|0\rangle.
\label{cpv}
\end{equation}
This means that if, in such a vacuum, none of the CP symmetries allowed by the Lagrangian satisfy Eq.~(\ref{cpv}), then the vacuum is not CP invariant; i.e., there is spontaneous CP violation.

\subsection{Complex vevs}
\label{Sec:complex}

We now study the complex vacua, i.e.
\begin{equation}
\label{cv1234theta}
(u_1,u_2,u_3,u_4)= (v_1 e^{i \theta_1}, v_2 e^{i \theta_2}, v_3 e^{i \theta_3}, v_4).
\end{equation}

$\bullet$ \textbf{ Normal vacua}

We now discuss the Normal vacua in the complex case, in which all $v_i$ are non-zero. Substituting Eq.~(\ref{cv1234theta}) into the stationarity conditions (\ref{Vu1})--(\ref{Vu4}) yields the following four expressions for $\mu^2_1$ and $\mu^2_2$:
\begin{equation}
\begin{aligned}
\label{cmu1v1}
   \mu^2_1=&\frac{1}{2}\left[2\lambda_1v^2_1 +\bar\lambda_{3}v^2_2+\bar\lambda_{5}v^2_3+\bar\lambda_{6}v^2_4+\bar\lambda_{11}\frac{v_2v_3v_4}{v_1}e^{i (\theta_3-\theta_1-\theta_2)}\right.\\
&\left.+\lambda_{13}\left(2v_2v_4e^{i (\theta_2-2\theta_1)}+\frac{v^2_2v_3}{v_1}e^{i( 2\theta_2-\theta_1-\theta_3)}\right)+\lambda_{14}\frac{v^2_3v_4}{v_1}e^{i(2\theta_3-\theta_1)}\right],
\end{aligned}
\end{equation}
\begin{equation}
\begin{aligned}
\label{cmu1v2}
   \mu^2_1=&\frac{1}{2}\left[2\lambda_1v^2_2 +\bar\lambda_{3}v^2_1+\bar\lambda_{5}v^2_4+\bar\lambda_{6}v^2_3+\bar\lambda_{11}\frac{v_1v_3v_4}{v_2}e^{i (\theta_3-\theta_1-\theta_2)}\right.\\
&\left.+\lambda_{13}\left(2v_1v_3e^{i (\theta_1+\theta_3-2\theta_2)}+\frac{v^2_1v_4}{v_2}e^{i(2\theta_1-\theta_2)}\right)+\lambda_{14}\frac{v_3v^2_4}{v_2}e^{-i(\theta_2+\theta_3)}\right],
\end{aligned}
\end{equation}
\begin{equation}
\begin{aligned}
\label{cmu2v3}
   \mu^2_2=&\frac{1}{2}\left[2\lambda_2v^2_3+\bar\lambda_{4}v^2_4+\bar\lambda_{5}v^2_1+\bar\lambda_{6}v^2_2+\bar\lambda_{11}\frac{v_1v_2v_4}{v_3}e^{i (\theta_1+\theta_2-\theta_3)}\right.\\
&\left.+\lambda_{13}\frac{v_1v^2_2}{v_3}e^{i(2\theta_2-\theta_1-\theta_3)}+\lambda_{14}\left(2v_1v_4e^{i (\theta_1-2\theta_3)}+\frac{v_2v^2_4}{v_3}e^{-i(\theta_2+\theta_3)}\right)\right],
\end{aligned}
\end{equation}
\begin{equation}
\begin{aligned}
\label{cmu2v4}
 \mu^2_2=&\frac{1}{2}\left [2\lambda_2v^2_4+\bar\lambda_{4}v^2_3+\bar\lambda_{5}v^2_2+\bar\lambda_{6}v^2_1+\bar\lambda_{11}\frac{v_1v_2v_3}{v_4}e^{i (\theta_1+\theta_2-\theta_3)}\right.\\
&\left.+\lambda_{13}\frac{v^2_1v_2}{v_4}e^{i (2\theta_1-\theta_2)}+\lambda_{14}\left(2v_2v_3e^{i(\theta_2+\theta_3)}+\frac{v_1v^2_3}{v_4}e^{i(2\theta_3-\theta_1)}\right)\right].
\end{aligned}
\end{equation}
Each of the four equations above contains both a real and an imaginary part; therefore, we have eight constraint equations. Analogous to the real vacua in Section~\ref{Sec:Real}, the four equations (\ref{cmu1v1}), (\ref{cmu1v2}), (\ref{cmu2v3}), and (\ref{cmu2v4}) are not applicable when $v_1 = 0$, $v_2 = 0$, $v_3 = 0$, and $v_4 = 0$, respectively. Since the two expressions for $\mu_1^2$ in (\ref{cmu1v1}) and (\ref{cmu1v2}) must be consistent with each other, and likewise for $\mu_2^2$ in (\ref{cmu2v3}) and (\ref{cmu2v4}), we obtain the following two consistency conditions:
\begin{subequations}
\begin{equation}
\begin{aligned}
\label{cmu1mmu1}
&\left[2\lambda_1-\bar\lambda_{3}-\bar\lambda_{11}\frac{v_3v_4}{v_1v_2}\cos(\theta_1+\theta_2-\theta_3)\right](v^2_1-v^2_2)+(\bar\lambda_{5}-\bar\lambda_{6})(v^2_3-v^2_4)\\
&+\lambda_{13}\left[\frac{v_3}{v_1}(v^2_2-2v^2_1)\cos(\theta_1+\theta_3-2\theta_2)+\frac{v_4}{v_2}(2v^2_2-v^2_1)\cos(2\theta_1-\theta_2)\right]\\
&+\lambda_{14}v_3v_4\left[\frac{v_3}{v_1}\cos(\theta_1-2\theta_3)-\frac{v_4}{v_2}\cos(\theta_2+\theta_3)\right]=0,
\end{aligned}
\end{equation}
\begin{equation}
\begin{aligned}
\label{cmu2mmu2}
&\left[2\lambda_2-\bar\lambda_{4}-\bar\lambda_{11}\frac{v_1v_2}{v_3v_4}\cos(\theta_1+\theta_2-\theta_3)\right](v^2_3-v^2_4)+(\bar\lambda_{5}-\bar\lambda_{6})(v^2_1-v^2_2)\\
&+\lambda_{13}v_1v_2\left[\frac{v_2}{v_3}\cos(\theta_1+\theta_3-2\theta_2)-\frac{v_1}{v_4}\cos(2\theta_1-\theta_2)\right]\\
&+\lambda_{14}\left[\frac{v_2}{v_3}(v^2_4-2v^2_3)\cos(\theta_2+\theta_3)+\frac{v_1}{v_4}(2v^2_4-v^2_3)\cos(\theta_1-2\theta_3)\right]=0.
\end{aligned}
\end{equation}
\end{subequations}
By simultaneously solving the equations obtained from the imaginary parts of Eqs.~(\ref{cmu1v1})--(\ref{cmu2v4}), we obtain the following three independent constraints:
\begin{subequations}
\begin{align}
&\lambda_{13}v_1v_2[v_2v_3\sin(\theta_1+\theta_3-2\theta_2)+v_1v_4\sin(2\theta_1-\theta_2)]=0,\label{lambda13}\\&
\lambda_{14}v_3v_4[v_2v_4\sin(\theta_2+\theta_3)-v_1v_3\sin(\theta_1-2\theta_3)]=0,\label{lambda14}\\&
v_1v_4[\bar\lambda_{11}v_2v_3\sin(\theta_1+\theta_2-\theta_3)+\lambda_{13}v_1v_2\sin(2\theta_1-\theta_2)+\lambda_{14}v^2_3\sin(\theta_1-2\theta_3)]=0.\label{lambda1112}
\end{align}
\end{subequations}
First, consider the most general complex vacuum ($v_1e^{i\theta_1}, v_2e^{i\theta_2}, v_3e^{i\theta_3}, v_4$), whose vevs comprise four arbitrary moduli $v_1, v_2, v_3, v_4$ and three arbitrary phases $\theta_1, \theta_2, \theta_3$. Solving Eqs.~(\ref{lambda13}) and (\ref{lambda14}), we find that this vacuum necessarily implies $\lambda_{13}=0$ and $\lambda_{14}=0$. Therefore, a vacuum in which both the parameters $v_i$ and $\theta_i$ can simultaneously take arbitrary values does not satisfy the conditions of the model. Further calculations show that only if the three phases $\theta_1$, $\theta_2$, and $\theta_3$ are constrained—i.e., they must satisfy a specific relation—can one avoid $\lambda_{13}=0$ and $\lambda_{14}=0$.

Imposing $\lambda_{13} \neq 0$ and $\lambda_{14} \neq 0$, we keep the four moduli $v_1, v_2, v_3, v_4$ arbitrary but lift the arbitrariness of the three phases $\theta_1, \theta_2, \theta_3$. From Eqs.~(\ref{lambda13}) and (\ref{lambda14}) we obtain: 
\begin{equation}  
  \sin(\theta_1+\theta_3-2\theta_2)=\sin(2\theta_1-\theta_2)=\sin(\theta_2+\theta_3)=\sin(\theta_1-2\theta_3)=0.
      \label{cv1234sin}
\end{equation}
Further solving yields the corresponding values of the three phases $\theta_1$, $\theta_2$, and $\theta_3$. Substituting these values into Eqs.~(\ref{cmu1v1})--(\ref{cmu2mmu2}) and (\ref{lambda1112}) gives four constraints on the potential parameters. This vacuum is denoted as C-N-4a in Table \ref{Table:cvev}.

Next, we consider cases in which some of the moduli $|v_i|$ are equal among the complex vevs. Under the conditions that $\lambda_{13} \neq 0$ and $\lambda_{14} \neq 0$, solving Eqs.~(\ref{cmu1mmu1}) and (\ref{cmu2mmu2}) yields the following solutions:
\begin{equation}
\label{cv1v1v3v3}
      \left |v_1 \right |=\left |v_2 \right |, \quad \text{and}\quad  \left |v_3 \right |=\left |v_4 \right |, 
\end{equation}
and
\begin{equation}  
     \cos(\theta_1+\theta_3-2\theta_2)=\cos(2\theta_1-\theta_2), \quad \text{and}\quad  \cos(\theta_2+\theta_3)=\cos(\theta_1-2\theta_3).
\label{ccos}
\end{equation}
Substituting Eq.~(\ref{cv1v1v3v3}) into Eqs.~(\ref{lambda13}) and (\ref{lambda14}) yields the following:
\begin{equation}  
    \sin(\theta_1+\theta_3-2\theta_2)=- \sin(2\theta_1-\theta_2), \quad \text{and}\quad  \sin(\theta_2+\theta_3)=\sin(\theta_1-2\theta_3).
      \label{csin}
\end{equation}
By combining Eqs.~(\ref{ccos}) and (\ref{csin}) and solving them, we obtain the values of the three phases $\theta_1$, $\theta_2$, and $\theta_3$. Substituting these phase values, together with Eq.~(\ref{cv1v1v3v3}), into Eqs.~(\ref{cmu1v1})--(\ref{cmu2v4}) and (\ref{lambda1112}) then yields constraints on the potential parameters. These vacua are denoted as C-N-2, C-N-3a, C-N-3b, C-N-3c, and C-N-3d in Table \ref{Table:cvev}.

$\bullet$ \textbf{Inert vacua}

We now discuss the Inert vacua among the complex vacua. First, we consider the case in which only one of the $v_i$ is zero. For example, taking $v_1=0$, the vevs are written as $(0, v_2e^{i\theta_2}, v_3e^{i\theta_3}, v_4)$. In this case, Eq.~(\ref{cmu1v1}) no longer applies; instead, we use the original stationarity condition (\ref{Vu1}). Furthermore, the expression for $\mu_1^2$ in (\ref{cmu1v2}) remains automatically consistent, whereas the two expressions for $\mu_2^2$, (\ref{cmu2v3}) and (\ref{cmu2v4}), must still satisfy the consistency condition (\ref{cmu2mmu2}). Assuming $\lambda_{13} \neq 0$ and $\lambda_{14} \neq 0$, solving these equations readily yields the values of $\theta_2$ and $\theta_3$ as: 
\begin{equation}
   \theta_2=\frac{n\pi}{5},\quad\text{and} \quad \theta_3=(-\frac{n}{5}+m)\pi,\quad m,n\text{ integer}.
\end{equation}
Since $e^{im\pi} = \pm 1$ and $v_2, v_3, v_4$ are arbitrary non-zero real values, we can absorb the factor $e^{im\pi}$ into the sign of $v_3$. Therefore, this vacuum can be written as $(0, v_2e^{\frac{in\pi}{5}}, v_3e^{-\frac{in\pi}{5}}, v_4)$, which is a solution of the model. For other complex vacua in which exactly one of the $v_i$ is zero, analogous configurations arise by permuting indices, and all satisfy the model requirements.

Second, we consider the case in which two of the $v_i$ are zero. This outcome is consistent with the results for the real vacua. For all complex vacua with two vanishing $v_i$, substituting them into the four stationarity conditions (\ref{Vu1})--(\ref{Vu4}) inevitably leads to at least one of $\lambda_{13}=0$ or $\lambda_{14}=0$. Thus, in this case, there is no vacuum that satisfies the model requirements.

Finally, we consider the case in which three of the $v_i$ are zero. Since a global rotation of the vacuum can always eliminate one phase, the vacuum is actually real in this case. This case has already been discussed in Section~\ref{Sec:Real} on real vacua.

In Table \ref{Table:cvev}, we summarize the complete list of complex vacua and explicitly indicate whether each vacuum exhibits spontaneous CP violation, marked as ``yes'' or ``no''. We also use the notation C-X-Yz to classify each complex vacuum. For vacua with a single unknown phase $\theta$ (originally $\theta_i$), we drop the subscript for brevity. Additionally, the constraints on the parameter space admitting each complex vacuum solution are summarized in Table \ref{Table:cvevcon}. 
\begin{table}[htb]
\centering
\begin{tabular}{|l|c|c|}
\hline 
Vacuum &Complex vevs &SCPV\\
\hline 
C-N-2&$v_1e^{\frac{in\pi}{5}},\pm v_1e^{\frac{2in\pi}{5}}, \pm v_3e^{\frac{3in\pi}{5}},v_3$  &no\\
\hline 
C-N-3a & $v_1e^{i\theta},\pm v_1e^{i(\theta+\frac{n\pi}{5})}, \pm v_3e^{\frac{3in\pi}{5}},v_3$ &yes\\
\hline 
C-N-3b & $v_1e^{i\theta},\pm v_1e^{i\theta},\pm v_3,v_3$ &yes\\
\hline 
C-N-3c&$v_1e^{i(\frac{1}{2}+\frac{n}{5})\pi},\pm v_1e^{i(\frac{1}{2}+\frac{2n}{5})\pi}, \pm v_3e^{\frac{3in\pi}{5}},v_3$   &yes\\
\hline 
C-N-3d&$iv_1, \pm iv_1,\pm v_3,v_3$   &yes\\
\hline 
C-N-4a&$v_1e^{\frac{2in\pi}{5}},v_2e^{-\frac{in\pi}{5}}, v_3e^{\frac{in\pi}{5}},v_4$  &no\\
\hline
C-N-4b&$v_1e^{\frac{2in\pi}{5}},\pm v_2e^{-\frac{in\pi}{5}}, \pm v_1e^{\frac{in\pi}{5}},v_2$  &no\\
\hline
C-N-4c&$v_1e^{\frac{2in\pi}{5}},\pm v_2e^{-\frac{in\pi}{5}}, v_2e^{\frac{in\pi}{5}},\pm v_1$  &no\\
\hline
C-N-4d&$v_1e^{\frac{2in\pi}{5}},\pm v_1e^{-\frac{in\pi}{5}}, \pm v_3e^{\frac{in\pi}{5}},v_3$  &no\\
\hline
C-I-4a & $v_1e^{\frac{2in\pi}{5}},v_2e^{\frac{in\pi }{5}},v_3, 0$&no \\
\hline 
C-I-4b & $ v_1e^{\frac{in\pi}{5}}, v_2e^{\frac{2in\pi}{5}}, 0,v_4$&no \\
\hline
C-I-4c & $v_1e^{\frac{2in\pi}{5}},0, v_3e^{\frac{in\pi}{5}}, v_4$&no \\
\hline
C-I-4d & $0,v_2e^{\frac{in\pi}{5}}, v_3e^{-\frac{in\pi}{5}}, v_4$&no \\
\hline
\end{tabular}
\caption{Complex vacua. In the notation C-X-Yz, C indicates that the vacuum is complex. The other rules are consistent with those for real vacua.
\label{Table:cvev}}
\end{table}
\begin{table}[htbp]
\centering
\begin{tabular}{|l|c|c|}
\hline 
Vacuum& Constraints  \\
\hline 
C-N-2&\makecell{$\mu^2_1=\frac{1}{2}\left[(2\lambda_1+\bar\lambda_{3})v^2_1+(\bar\lambda_{5}+\bar\lambda_{6}+\bar\lambda_{11})v^2_3\pm3\lambda_{13}v_1v_3+(-1)^n\lambda_{14}\frac{v^3_3}{v_1}\right]$,\\$\mu^2_2=\frac{1}{2}\left[(2\lambda_2+\bar\lambda_{4})v^2_3+(\bar\lambda_{5}+\bar\lambda_{6}+\bar\lambda_{11})v^2_1\pm\lambda_{13}\frac{v^3_1}{v_3}+(-1)^n3\lambda_{14}v_1v_3\right]$}\\
\hline 
C-N-3a &\makecell{$\mu^2_1=\frac{1}{2}\left[\left(2\lambda_1+\bar\lambda_{3}\right)v^2_1+\left(\bar\lambda_{5}+\bar\lambda_{6}+\bar\lambda_{11}\cos2\left(\theta-\frac{n\pi}{5}\right)\right)v^2_3\right.$\\$\left.\pm3\lambda_{13}v_1v_3\cos\left(\theta-\frac{n\pi}{5}\right)+\lambda_{14}\frac{v^3_3}{v_1}\cos\left(\theta-\frac{6n\pi}{5}\right)\right]$,\\$\mu^2_2=\frac{1}{2}\left[\left(2\lambda_2+\bar\lambda_{4}\right)v^2_3+\left(\bar\lambda_{5}+\bar\lambda_{6}+\bar\lambda_{11}\cos2\left(\theta-\frac{n\pi}{5}\right)\right)v^2_1\right.$\\$\left.\pm\lambda_{13}\frac{v^3_1}{v_3}\cos\left(\theta-\frac{n\pi}{5}\right)+3\lambda_{14}v_1v_3\cos\left(\theta-\frac{6n\pi}{5}\right)\right]$,\\$\left[\lambda_{13}v_1^2\pm(-1)^n\lambda_{14}v_3^2\pm2\bar\lambda_{11}v_1v_3\cos\left(\theta-\frac{n\pi}{5}\right)\right]\sin\left(\theta-\frac{n\pi}{5}\right)=0$}\\
\hline 
C-N-3b &\makecell{$\mu^2_1=\frac{1}{2}\left[(2\lambda_1+\bar\lambda_{3})v^2_1+(\bar\lambda_{5}+\bar\lambda_{6}+\bar\lambda_{11}\cos2\theta)v^2_3+\left(\lambda_{14}\frac{v^3_3}{v_1}\pm3\lambda_{13}v_1v_3\right)\cos\theta\right]$,\\$\mu^2_2=\frac{1}{2}\left[(2\lambda_2+\bar\lambda_{4})v^2_3+(\bar\lambda_{5}+\bar\lambda_{6}+\bar\lambda_{11}\cos2\theta)v^2_1+\left(3\lambda_{14}v_1v_3\pm\lambda_{13}\frac{v^3_1}{v_3}\right)\cos\theta\right],$\\$\lambda_{13}=\mp\frac{\lambda_{14}v^2_3+2\bar\lambda_{11}v_1v_3\cos\theta}{v^2_1}$}\\
\hline 
\makecell{ C-N-3c,\\C-N-3d} &\makecell{$\mu^2_1=\frac{1}{2}\left[(2\lambda_1+\bar\lambda_{3})v^2_1+(\bar\lambda_{5}+\bar\lambda_{6}-\bar\lambda_{11})\right]v^2_3$,\\$\mu^2_2=\frac{1}{2}\left[(2\lambda_2+\bar\lambda_{4})v^2_3+(\bar\lambda_{5}+\bar\lambda_{6}-\bar\lambda_{11})\right]v^2_1,\lambda_{13}=\mp (-1)^n\lambda_{14}\frac{v^2_3}{v^2_1}$}\\
\hline
C-N-4a  &Eqs.~(\ref{cn4ac1}), (\ref{cn4ac2}), (\ref{cn4ac3}), (\ref{cn4ac4})\\
\hline 
C-N-4b  &\makecell[c]{$\mu^2_1=\frac{1}{2}[(2\lambda_1+\bar\lambda_{5})(v_1^2+v_2^2)+\lambda_{14}v_1v_2],$\\$\mu^2_2=\frac{1}{2}[(2\lambda_2+\bar\lambda_{5})(v_1^2+v_2^2)+\lambda_{14}v_1v_2]$,\\$\lambda_1=\frac{1}{2}(\bar\lambda_{3}-\bar\lambda_{5}+\bar\lambda_{6}+\bar\lambda_{11}\pm(-1)^n3\lambda_{13})$,\\$\lambda_2=\frac{1}{2}\left(\bar\lambda_{4}-\bar\lambda_{5}+\bar\lambda_{6}+\bar\lambda_{11}\pm (-1)^n\lambda_{13}+\lambda_{14}\frac{v_1^2+v_2^2}{v_1v_2}\right)$}\\
\hline 
C-N-4c  &\makecell[c]{$\mu^2_1=\frac{1}{2}[(2\lambda_1+\bar\lambda_{6})(v_1^2+v_2^2)+\lambda_{13}v_1v_2],$\\$\mu^2_2=\frac{1}{2}[(2\lambda_2+\bar\lambda_{6})(v_1^2+v_2^2)+\lambda_{13}v_1v_2]$,\\$\lambda_1=\frac{1}{2}\left(\bar\lambda_{3}+\bar\lambda_{5}-\bar\lambda_{6}+\bar\lambda_{11}\pm\lambda_{14}+(-1)^n\lambda_{13}\frac{v_1^2+v_2^2}{v_1v_2}\right)$,\\$\lambda_2=\frac{1}{2}\left(\bar\lambda_{4}+\bar\lambda_{5}-\bar\lambda_{6}+\bar\lambda_{11}\pm3\lambda_{14}\right)$}\\
\hline 
C-N-4d&\makecell{$\mu^2_1=\frac{1}{2}\left[(2\lambda_1+\bar\lambda_{3})v^2_1+(\bar\lambda_{5}+\bar\lambda_{6}+\bar\lambda_{11})v^2_3\pm(-1)^n3\lambda_{13}v_1v_3+\lambda_{14}\frac{v^3_3}{v_1}\right]$,\\$\mu^2_2=\frac{1}{2}\left[(2\lambda_2+\bar\lambda_{4})v^2_3+(\bar\lambda_{5}+\bar\lambda_{6}+\bar\lambda_{11})v^2_1\pm (-1)^n\lambda_{13}\frac{v^3_1}{v_3}+3\lambda_{14}v_1v_3\right]$}\\
\hline 
C-I-4a &\makecell[c]{$ 
\mu^2_1=\lambda_1v_2^2+\frac{1}{2}(\bar\lambda_{3}v_1^2+\bar\lambda_{6}v^2_3)+\lambda_{13}v_1v_3,\lambda_{14}=-\frac{(-1)^n(\bar\lambda_{11}v_3+\lambda_{13}v_1)v_2}{v_3^2},$\\$\mu^2_2=\lambda_2v_3^2+\frac{1}{2}\left(\bar\lambda_{5}v_1^2+\bar\lambda_{6}v^2_2+\lambda_{13}\frac{v_1v_2^2}{v_3}\right),$\\$ \lambda_{13}\frac{v_3(2v_1^2-v_2^2)}{v_1}=(2\lambda_1-\bar\lambda_{3})(v_1^2-v_2^2)+(\bar\lambda_{5}-\bar\lambda_{6})v_3^2,$} \\
\hline 
C-I-4b &\makecell[c]{$ 
\mu^2_1=\lambda_1v_1^2+\frac{1}{2}(\bar\lambda_{3}v_2^2+\bar\lambda_{6}v^2_4)+\lambda_{13}v_2v_4,\lambda_{14}=-\frac{(-1)^n(\bar\lambda_{11}v_4+\lambda_{13}v_2)v_1}{v_4^2},$\\$\mu^2_2=\lambda_2v_4^2+\frac{1}{2}\left(\bar\lambda_{5}v_2^2+\bar\lambda_{6}v^2_1+\lambda_{13}\frac{v_1^2v_2}{v_4}\right),$\\$ \lambda_{13}\frac{v_4(2v_2^2-v_1^2)}{v_2}=(2\lambda_1-\bar\lambda_{3})(v_2^2-v_1^2)+(\bar\lambda_{5}-\bar\lambda_{6})v_4^2,$} \\
\hline 
C-I-4c&\makecell[c]{ $\mu^2_1=\lambda_1v_1^2+\frac{1}{2}\left(\bar\lambda_{5}v_3^2+\bar\lambda_{6}v_4^2+\lambda_{14}\frac{v_3^2v_4}{v_1}\right),$\\$ 
\mu^2_2=\lambda_2v_3^2+\frac{1}{2}(\bar\lambda_{4}v_4^2+\bar\lambda_{5}v^2_1)+\lambda_{14}v_1v_4, \lambda_{13}=-\frac{(-1)^n(\bar\lambda_{11}v_1+\lambda_{14}v_4)v_3}{v_1^2} ,$\\$\lambda_{14}\frac{v_1(2v_4^2-v_3^2)}{v_4}=(2\lambda_2-\bar\lambda_{4})(v_4^2-v_3^2)+(\bar\lambda_{6}-\bar\lambda_{5})v_1^2$} \\
\hline 
C-I-4d &\makecell[c]{$\mu^2_1=\lambda_1v_2^2+\frac{1}{2}\left(\bar\lambda_{5}v_4^2+\bar\lambda_{6}v^2_3+\lambda_{14}\frac{v_3v_4^2}{v_2}\right),$\\$\mu^2_2=\lambda_2v_4^2+\frac{1}{2}(\bar\lambda_{4}v_3^2+\bar\lambda_{5}v_2^2)+\lambda_{14}v_2v_3, \lambda_{13}=-\frac{(-1)^n(\bar\lambda_{11}v_2+\lambda_{14}v_3)v_4}{v^2_2}$,\\$\lambda_{14}\frac{v_2(2v_3^2-v_4^2)}{v_3}=(2\lambda_2-\bar\lambda_{4})(v_3^2-v_4^2)+(\bar\lambda_{6}-\bar\lambda_{5})v_2^2$} \\
\hline
\end{tabular}
\caption{Constraints on complex vacua.
\label{Table:cvevcon}}
\end{table}

We now discuss each complex vacuum listed in Table \ref{Table:cvev} and explain why it does or does not lead to spontaneous CP violation.
\begin{itemize}
	\item
The vacuum C-N-2 satisfies $v_1^2 + v_3^2 = \frac{v^2_{w}}{2}$. When the integer $n$ is divisible by 5 (i.e., $5 \mid n$), the phases become trivial and this vacuum becomes the real vacuum R-N-2. When $n$ is not divisible by 5 (i.e., $5 \nmid n$), both forms of this vacuum completely break the $D_5$ symmetry and exhibit three calculable, nontrivial phases. These phases are entirely determined by the $D_5$ symmetry of the potential, are independent of its coupling constants, and cannot be rotated away. Because there exist two matrices $U_1$ and $U_2$ that satisfy the constraints of Eq.~(\ref{cpv}), namely,
\begin{equation}
U_1 = \operatorname{diag}\left(\eta,\eta^2,\eta^3,1\right),\eta=e^{\frac{2in\pi}{5}},\quad \quad
U_2 =\pm e^{\frac{3in\pi}{5}} \left( \begin{array}{cccc}
0 & 1 & 0 & 0\\
1 & 0 & 0 & 0\\
0 & 0 & 0 & 1\\
0 & 0 & 1 & 0
\end{array} \right).
\label{cpcU1U2}
\end{equation}
Since either one alone suffices to preclude spontaneous CP violation, there is no spontaneous CP violation in this vacuum. Moreover, $U_2$ is also a symmetry of the potential. This indicates that a complex vacuum does not necessarily lead to spontaneous CP violation.
  \item
Vacuum C-N-3a satisfies $v_1^2 + v_3^2 = \frac{v^2_{w}}{2}$. When the integer $n$ is divisible by $5$ (i.e., $5 \mid n$), it reduces to Vacuum C-N-3b, leaving a residual $Z_2$ symmetry. When $n$ is not divisible by 5 (i.e., $5 \nmid n$), both forms of this vacuum completely break the $D_5$ symmetry. This vacuum allows for three non-trivial phases. Two of them can be determined as functions of $\bar\lambda_{11}$, $\lambda_{13}$, and $\lambda_{14}$, while the third one is independent of the coupling constants, as shown in Table \ref{Table:cvevcon}. We have found that when $\theta$ satisfies
\begin{equation}
 \theta\neq\frac{n\pi}{5}, \quad n\text{ integer}. 
\label{thetaneqn5}
\end{equation}
This vacuum exhibits spontaneous CP violation. If $\theta=\frac{n\pi}{5}$, Vacuum C-N-3a reduces to Vacuum C-N-2, and spontaneous CP violation no longer occurs. Condition (\ref{thetaneqn5}) is necessary and sufficient for C-N-3a to exhibit spontaneous CP violation.
 \item
 Vacuum C-N-3b can be viewed as a special case of C-N-3a when $n=0$. It contains $(v_1e^{i\theta}, v_1e^{i\theta}, v_3, v_3)$ and $(v_1e^{i\theta}, -v_1e^{i\theta}, -v_3, v_3)$. The former preserves a residual $Z_2$ symmetry, whereas the latter completely breaks the $D_5$ symmetry. This vacuum admits two identical non-trivial phases, which can be determined as functions of $\bar\lambda_{11}$, $\lambda_{13}$, and $\lambda_{14}$, as shown in Table \ref{Table:cvevcon}. We have found that when $\theta$ satisfies
\begin{equation}
\theta\neq k\pi ,\quad  k\text{ integer}.
\label{thetaneqk}
\end{equation}   
This vacuum violates CP spontaneously.
\item
Vacuum C-N-3c satisfies $v_1^2 + v_3^2 = \frac{v^2_{w}}{2}$. When $n=0$, it reduces to Vacuum C-N-3d, which leaves a residual $Z_2$ symmetry. When $n$ is a non-zero integer, both forms of this vacuum completely break the $D_5$ symmetry and exhibit three calculable nontrivial phases. As in the case of C-N-2, these phases are entirely determined by the $D_5$ symmetry of the potential, are independent of its coupling constants, and cannot be rotated away. However, this vacuum violates CP spontaneously. Moreover, Vacuum C-N-3c can be seen as a special case of C-N-3a when $\theta= (\frac{1}{2}+\frac{n}{5})\pi$.
  \item
Vacuum C-N-3d can be seen as a special case of C-N-3c when $n=0$. It contains $(iv_1, iv_1, v_3, v_3)$ and $(iv_1, -iv_1, -v_3, v_3)$. The former leaves a residual $Z_2$ symmetry, while the latter completely breaks the $D_5$ symmetry. This vacuum exhibits two identical, calculable, nontrivial phases, which are entirely determined by the $D_5$ symmetry of the potential, are independent of its coupling constants, and cannot be rotated away. This vacuum violates CP spontaneously.
 \item
Vacuum C-N-4a completely breaks the $D_5$ symmetry. When $5 \mid n$, its phases become trivial and this vacuum becomes the real vacuum R-N-4a. When $5 \nmid n$, this vacuum exhibits three calculable nontrivial phases, which are entirely determined by the $D_5$ symmetry of the potential, are independent of its coupling constants, and cannot be rotated away. Due to the existence of a matrix $U$ that satisfies the constraint of Eq.~(\ref{cpv}), namely
\begin{equation}
U = \operatorname{diag}(\eta^2,\eta^4,\eta, 1),\quad \eta=e^{\frac{2in\pi}{5}}.
\label{cpvu}
\end{equation}
There is no spontaneous CP violation in this vacuum.
\item
The vacua C-N-4b and C-N-4c are related by a permutation of the Higgs fields, satisfy $v_1^2 + v_2^2 = \frac{v^2_{w}}{2}$, and completely break $D_5$. They can be seen as special cases of C-N-4a where some of the moduli $|v_i|$ become equal. Since there exists a matrix $U$ (given in Eq.~(\ref{cpvu})) that satisfies Eq.~(\ref{cpv}), CP is not spontaneously violated in these vacua either.
\item
Vacuum C-N-4d can also be seen as a special case of C-N-4a where some of the moduli $|v_i|$ become equal, but it actually has only two constraints on the potential parameters. It is related to C-N-2 by a permutation of the Higgs fields, and the two vacua impose almost the same constraints. To emphasize that it can be obtained from C-N-4a and to distinguish it from C-N-2, the notation of this vacuum does not reflect the number of constraints. C-N-4d can be analyzed by reasoning analogous to that for C-N-2. Due to the existence of two matrices $U$ (given in Eq.~(\ref{cpvu})) and $U'$ that satisfy Eq.~(\ref{cpv}), namely
\begin{equation}
U' =\pm e^{\frac{in\pi}{5}} \left( \begin{array}{cccc}
0 & 1 & 0 & 0\\
1 & 0 & 0 & 0\\
0 & 0 & 0 & 1\\
0 & 0 & 1 & 0
\end{array} \right).
\label{cpvU'}
\end{equation}
Since either one alone suffices to preclude spontaneous CP violation, there is no spontaneous CP violation in this vacuum. Moreover, $U'$ is also a symmetry of the potential.

\item
Vacua C-I-4a, C-I-4b, C-I-4c, and C-I-4d completely break the $D_5$ symmetry. When $5 \mid n$, their phases become trivial, and these vacua reduce to the real vacua R-I-4a, R-I-4b, R-I-4c, and R-I-4d, respectively. When $5 \nmid n$, each such vacuum exhibits two calculable nontrivial phases, which are entirely determined by the $D_5$ symmetry of the potential, independent of its coupling constants, and cannot be rotated away. However, these four vacua do not lead to spontaneous CP violation due to the existence of the matrices $U_1 = \operatorname{diag}\left(\eta^2,\eta,1,\eta^3\right)$, $
U_2 = \operatorname{diag}\left(\eta,\eta^2,\eta^3,1\right)$, $
U_3 = \operatorname{diag}\left(\eta^2,\eta^4,\eta,1\right)$, and $
U_4 = \operatorname{diag}\left(\eta^3,\eta,\eta^4,1\right)$, respectively, where $\eta=e^{\frac{2in\pi}{5}}$, each of which satisfies the constraint of Eq.~(\ref{cpv}).

\end{itemize}

\paragraph{Relationships among complex vacua} 

As summarized in Table \ref{Table:cvev}, all complex vacua can be further classified into three types: 
\begin{itemize}
\item[(1)] The first type consists of the Normal vacua with SCPV: C-N-3a, C-N-3b, C-N-3c, and C-N-3d. Vacuum C-N-3a can be regarded as a more general case of the other three vacua, since each can be obtained by assigning specific values to its parameters $v_i$, $\theta$, and $n$.

\item[(2)] The second type consists of the Normal vacua without SCPV: C-N-2, C-N-4a, C-N-4b, C-N-4c, and C-N-4d. Vacuum C-N-4a can be regarded as a more general case encompassing C-N-4b, C-N-4c, and C-N-4d, since each can be obtained by assigning specific values to its parameters $v_i$.

\item[(3)] The third type consists of the Inert vacua without SCPV: C-I-4a, C-I-4b, C-I-4c, and C-I-4d.
\end{itemize}

To be local minima of the scalar potential, these complex vacua must satisfy the positive-definiteness conditions for their Hessians (\ref{Hessianc}). A detailed discussion is provided in Appendix~C.

\section{Residual $Z_2$ symmetry}
\label{Sec:Z2}

Among the vacua summarized in this work, three vevs preserve a residual $Z_2$ symmetry of the scalar potential after SSB:
\begin{equation}
\begin{aligned}
&(v_1,v_1,v_3,v_3),\quad (v_1e^{i\theta}, v_1e^{i\theta}, v_3, v_3),\quad(iv_1, iv_1, v_3, v_3). 
\end{aligned}
\end{equation}
It was shown in Ref.~\cite{GonzalezFelipe:2014mcf} that in an N-Higgs-doublet model, if the vev leaves a nontrivial residual symmetry (apart from the baryon number symmetry $U(1)_B$), this leads to unphysical quark masses and an unphysical CKM matrix. The consequences include: (i) a block-diagonal CKM matrix; (ii) degenerate quark masses; (iii) massless quarks. Therefore, to avoid such unphysical quark masses and mixings, the vev must completely break the flavour symmetry of the potential. 

One way to proceed is to add new interaction terms to the scalar potential that explicitly break $D_5$ symmetry. In this case, the model is no longer an exactly $D_5$-symmetric 4HDM, but rather a modified model. Consequently, the stationarity conditions and the corresponding vacuum solutions will also change, thereby affecting the complex vacua that might lead to spontaneous CP violation. We plan to investigate this type of model in future work.

\section{Complex vacua vs real vacua}
\label{Sec:Complexvsreal}

In our $D_5$ 4HDM, complex vacua are characterized by four moduli and three phases; their vevs are obtained by solving seven stationarity conditions. In contrast, real vacua are characterized by four moduli and satisfy three stationarity conditions. In the discussion of complex vacua in Section~\ref{Sec:complex}, when $5\mid n$, some complex vacua become real. We are interested in the relationship between these two types of vacua. In addition, a key question arises: which vacuum is deeper—the complex or the real?

Following Ref.~\cite{Emmanuel-Costa:2016vej}, let $\mathcal{C}(\text{C-X-Yz})$ denote the set of constraints (see Table \ref{Table:cvevcon}) satisfied by a specific complex vacuum, and let $\mathcal{C}(\text{R-X}' \text{-Y}'\text{z}')$ denote the set of constraints (see Table \ref{Table:rvev}) satisfied by a specific real vacuum. Then, if there exists a real limit of the vacuum (possibly up to a sign) in which C-X-Yz coincides in specification with R-X$'$-Y$'$z$'$, and the constraints are compatible, i.e., $ \mathcal{C}(\text{C-X-Yz})\subset  \mathcal{C}(\text{R-X}' \text{-Y}'\text{z}')$, we may regard the real vacuum R-X$'$-Y$'$z$'$ as the “origin” of the specific complex vacuum C-X-Yz. We list the real vacua that satisfy these requirements in Table~\ref{cvsr}.
\begin{table}[htbp]
\centering
\begin{tabular}{|c|c|}
\hline 
\makecell[c]{Complex } &Real ``origin"\\
\hline 
 C-N-2,3a,3b&R-N-2\\
\hline 
 C-N-3c,3d &none\\
\hline 
 C-N-4a&R-N-4a \\
\hline
 C-N-4b&R-N-4b \\
\hline
 C-N-4c&R-N-4c \\
\hline
 C-N-4d&R-N-2 \\
 \hline
 C-I-4a&R-I-4a \\
\hline 
 C-I-4b&R-I-4b \\
\hline 
 C-I-4c&R-I-4c \\
\hline 
 C-I-4d&R-I-4d \\
\hline 
\end{tabular}
\caption{Transitions from complex to real vacua.\label{cvsr}}
\end{table}

Except for Vacua C-N-3c and C-N-3d, which cannot be transformed into real vacua, all other complex vacua can be regarded as complex generalizations of corresponding real vacua. There are two main differences between real and complex vacua. First, the number of constraints for complex vacua is generally no less than that for real vacua, implying tighter restrictions on certain potential parameters in the complex case. Second, the Hessian’s positive-definiteness conditions differ: a real vacuum requires its four leading principal minors to be positive, whereas a complex vacuum requires seven. Given their more intricate structure, we provide explicit expressions only for the first four minors; the remaining ones must be verified numerically or by other means. In summary, complex vacua involve more phase degrees of freedom and are therefore subject to significantly more restrictive constraints than real vacua.

To compare the potential depths of two vacua, one must first determine whether they can coexist as local minima. This requires identifying a region in parameter space where both the potential constraints and the positive-definiteness conditions are satisfied simultaneously. In the $D_5$ 4HDM, establishing such coexistence is a formidable challenge, given the complexity of the vacuum structure, the large number of parameters, and the intricacy of the positive-definiteness conditions. If such a parameter region exists, the two vacua can both be local minima, and one may compare their depths by evaluating \(\Delta V = V_{\text{R}} - V_{\text{C}}\), which generally depends on the chosen parameters. Comparing real and complex vacua is a prerequisite for determining the global vacuum minimum, which is crucial for subsequent phenomenological analyses.

\section{Conclusion}
\label{Sec:Conclusion}

In this paper, we construct a four-Higgs-doublet model (4HDM) based on $D_5$ symmetry, which exhibits extremely rich phenomenology. Its scalar potential is described by 16 parameters, and the physical spectrum consists of 3 pairs of charged Higgs bosons and 7 neutral Higgs bosons, corresponding to 13 physical scalars after the 3 Goldstone modes are absorbed by the $W^\pm$ and $Z$ bosons. In addition, the model offers possibilities for CP violation and baryogenesis.

The $D_5$ 4HDM also possesses a rich vacuum structure. Assuming explicit CP conservation in the scalar potential, we comprehensively analyze its full neutral vacuum structure, systematically identify all possible real and complex vacuum solutions, and study the conditions for them to be local minima of the potential, namely the stationarity conditions and the positive-definiteness of the Hessian. In contrast to previous studies that mostly focused on real vacua, this work emphasizes complex vacua, aiming to assess whether the CP symmetry of the scalar potential can be spontaneously broken. We demonstrate that among the identified complex vacuum solutions (as shown in Table \ref{Table:cvev}), several specific complex vacua can spontaneously break CP symmetry. Among them, Vacuum C-N-3a, i.e., $(v_1e^{i\theta},\pm v_1e^{i(\theta+\frac{n\pi}{5})}, \pm v_3e^{\frac{3in\pi}{5}},v_3)$, is the most general complex vacuum that can lead to spontaneous CP violation. Moreover, comparing the real and complex vacua reveals that the latter can be regarded as complex generalizations of certain real vacua and impose more restrictive constraints on the potential parameters.

In summary, the $D_5$ 4HDM provides a natural and viable possibility for CP violation in the study of physics beyond the Standard Model, and our detailed analysis of its neutral vacuum structure lays a solid foundation for subsequent research. This work primarily focuses on the scalar sector, and future efforts will be dedicated to applying these vacua to the Yukawa sector to investigate fermion masses and mixing, thereby achieving a complete construction of the model.

\section*{Acknowledgments}
This work is supported by the Fundamental Research Funds for the Central Universities, the One Hundred Talent Program of Sun Yat-sen University, China, and the Guangdong Natural Science Foundation (Project No. 2026A1515012641).

\appendix

\section{Constraints of Vacuum C-N-4a}
\label{Sec:VacuumCN4a}

For Vacuum C-N-4a, the following four constraints apply:
\begin{subequations}
\begin{equation}
\label{cn4ac1}
   \mu^2_1=\frac{1}{2}\left[2\lambda_1v^2_1 +\bar\lambda_{3}v^2_2+\bar\lambda_{5}v^2_3+\bar\lambda_{6}v^2_4+\bar\lambda_{11}\frac{v_2v_3v_4}{v_1}+(-1)^n\lambda_{13}\left(2v_2v_4+\frac{v^2_2v_3}{v_1}\right)+\lambda_{14}\frac{v^2_3v_4}{v_1}\right],
\end{equation}
\begin{equation}
\label{cn4ac2}
   \mu^2_2=\frac{1}{2}\left[2\lambda_2v^2_3+\bar\lambda_{4}v^2_4+\bar\lambda_{5}v^2_1+\bar\lambda_{6}v^2_2+\bar\lambda_{11}\frac{v_1v_2v_4}{v_3}+(-1)^n\lambda_{13}\frac{v_1v^2_2}{v_3}+\lambda_{14}\left(2v_1v_4+\frac{v_2v^2_4}{v_3}\right)\right],
\end{equation}
\begin{equation}
\begin{aligned}
\label{cn4ac3}
&\left(2\lambda_1-\bar\lambda_{3}-\bar\lambda_{11}\frac{v_3v_4}{v_1v_2}\right)(v^2_1-v^2_2)+\left(\bar\lambda_{5}-\bar\lambda_{6}\right)(v^2_3-v^2_4)\\
&+(-1)^n\lambda_{13}\left(\frac{v_3(v_2^2-2v_1^2)}{v_1}+\frac{v_4(2v^2_2-v_1^2)}{v_2}\right)+\lambda_{14}v_3v_4\left(\frac{v_3}{v_1}-\frac{v_4}{v_2}\right)=0,
\end{aligned}
\end{equation}
\begin{equation}
\begin{aligned}
\label{cn4ac4}
&\left(2\lambda_2-\bar\lambda_{4}-\bar\lambda_{11}\frac{v_1v_2}{v_3v_4}\right)(v^2_3-v^2_4)+(\bar\lambda_{5}-\bar\lambda_{6})(v^2_1-v^2_2)\\
&+(-1)^n\lambda_{13}v_1v_2\left(\frac{v_2}{v_3}-\frac{v_1}{v_4}\right)+\lambda_{14}\left(\frac{v_2(v^2_4-2v_3^2)}{v_3}+\frac{v_1(2v_4^2-v_3^2)}{v_4}\right)=0.
\end{aligned}
\end{equation}
\end{subequations}

\section{Stationarity conditions in terms of moduli and phases}
\label{Sec:moduliphases}

The stationarity conditions for the moduli in Eq.~(\ref{vtheta}) can be written as:
\begin{equation}
\begin{aligned}
\frac{\partial V}{\partial v_1}=&-\mu^2_1v_1+\lambda_1v_1^3+\frac{1}{2}\left[v_1(\bar\lambda_3v_2^2+\bar\lambda_5v_3^2+\bar\lambda_6v_4^2)+\bar\lambda_{11}v_2v_3v_4 \cos(\theta_1+\theta_2-\theta_3)\right.\\
&\left.+\lambda_{13}\left(2v_1v_2v_4\cos(2\theta_1-\theta_2)+v_3v_2^2\cos(\theta_1+\theta_3-2\theta_2)\right)+\lambda_{14}v_4v_3^2\cos(\theta_1-2\theta_3)\right]=0,
\end{aligned}
\end{equation}
\begin{equation}
\begin{aligned}
\frac{\partial V}{\partial v_2}=&-\mu^2_1v_2+\lambda_1v_2^3+\frac{1}{2}\left[v_2(\bar\lambda_3v_1^2+\bar\lambda_5v_4^2+\bar\lambda_6v_3^2)+\bar\lambda_{11}v_1v_3v_4 \cos(\theta_1+\theta_2-\theta_3)\right.\\
&\left.+\lambda_{13}\left(2v_1v_2v_3\cos(\theta_1+\theta_3-2\theta_2)+v_1^2v_4\cos(2\theta_1-\theta_2)\right)+\lambda_{14}v_3v_4^2\cos(\theta_2+\theta_3)\right]=0,
\end{aligned}
\end{equation}
\begin{equation}
\begin{aligned}
\frac{\partial V}{\partial v_3}=&-\mu^2_2v_3+\lambda_2v_3^3+\frac{1}{2}\left[v_3(\bar\lambda_4v_4^2+\bar\lambda_5v_1^2+\bar\lambda_6v_2^2)+\bar\lambda_{11}v_1v_2v_4 \cos(\theta_1+\theta_2-\theta_3)\right.\\
&\left.+\lambda_{13}v_1v^2_2\cos(\theta_1+\theta_3-2\theta_2)+\lambda_{14}\left(2v_1v_3v_4\cos(\theta_1-2\theta_3)+v_2v_4^2\cos(\theta_2+\theta_3)\right)\right]=0,
\end{aligned}
\end{equation}
\begin{equation}
\begin{aligned}
\frac{\partial V}{\partial v_4}=&-\mu^2_2v_4+\lambda_2v_4^3+\frac{1}{2}\left[v_4(\bar\lambda_4v_3^2+\bar\lambda_5v_2^2+\bar\lambda_6v_1^2)+\bar\lambda_{11}v_1v_2v_3 \cos(\theta_1+\theta_2-\theta_3)\right.\\
&\left.+\lambda_{13}v_1^2v_2\cos(2\theta_1-\theta_2)+\lambda_{14}\left(2v_2v_3v_4\cos(\theta_2+\theta_3)+v_1v_3^2\cos(\theta_1-2\theta_3)\right)\right]=0,
\end{aligned}
\end{equation}
The stationarity conditions for the phases can be written as
\begin{equation}
\begin{aligned}
\label{theta1}
\frac{\partial V}{\partial \theta_1}=&-\frac{1}{2}\left[\bar\lambda_{11}v_1v_2v_3v_4 \sin(\theta_1+\theta_2-\theta_3)+\lambda_{14}v_1v_3^2v_4\sin(\theta_1-2\theta_3)\right.\\&\left.+\lambda_{13}\left(2v^2_1v_2v_4\sin(2\theta_1-\theta_2)+v_1v_2^2v_3\sin(\theta_1+\theta_3-2\theta_2)\right)\right]=0,
\end{aligned}
\end{equation}
\begin{equation}
\begin{aligned}
\label{theta2}
   \frac{\partial V}{\partial \theta_2}=&-\frac{1}{2}\left[\bar\lambda_{11}v_1v_2v_3v_4 \sin(\theta_1+\theta_2-\theta_3)+\lambda_{14}v_2v_3v^2_4\sin(\theta_2+\theta_3)\right.\\&\left.-\lambda_{13}\left(v^2_1v_2v_4\sin(2\theta_1-\theta_2)+2v_1v^2_2v_3\sin(\theta_1+\theta_3-2\theta_2)\right)\right]=0,
\end{aligned}
\end{equation}
\begin{equation}
\begin{aligned}
\label{theta3}
   \frac{\partial V}{\partial \theta_3}=&\frac{1}{2}\left[\bar\lambda_{11}v_1v_2v_3v_4 \sin(\theta_1+\theta_2-\theta_3)-\lambda_{13}v_1v^2_2v_3\sin(\theta_1+\theta_3-2\theta_2)\right.\\&\left.+\lambda_{14}\left(2v_1v_3^2v_4\sin(\theta_1-2\theta_3)-v_2v_3v^2_4\sin(\theta_2+\theta_3)\right)\right]=0.
\end{aligned}
\end{equation}

\section{Positive definiteness of the Hessian}
\label{Sec:Hessian}
\subsection{Real vacua}
\label{Hessianreal}

For R-N-4a, positive definiteness requires:
\begin{equation}
\begin{aligned}
&\bullet \quad D_1=a_{11}>0,\\
&\bullet \quad D_2= a_{11} a_{22} -a_{12}^2 >0,\\
&\bullet \quad D_3=a_{11}\left(a_{22} a_{33} - a_{23}^2\right) - a_{12}^2 a_{33} + 2 a_{12} a_{13} a_{23} - a_{13}^2 a_{22}>0,\\
&\bullet \quad D_4=a_{11} \left(a_{22} a_{33} a_{44} - a_{22} a_{34}^2 - a_{23}^2 a_{44} + 2 a_{23} a_{24} a_{34} - a_{24}^2 a_{33}\right)\\
  &\quad \quad- a_{12} \left(a_{12} a_{33} a_{44} - a_{12} a_{34}^2 - a_{13} a_{23} a_{44} + a_{13} a_{24} a_{34} + a_{14} a_{23} a_{34} - a_{14} a_{24} a_{33}\right)\\
  &\quad \quad+ a_{13} \left(a_{12} a_{23} a_{44} - a_{12} a_{24} a_{34} - a_{22} a_{13} a_{44} + a_{22} a_{14} a_{34} + a_{13} a_{24}^2 - a_{14} a_{24} a_{23}\right)\\
  &\quad \quad- a_{14} \left(a_{12} a_{23} a_{34} - a_{12} a_{24} a_{33} - a_{22} a_{13} a_{34} + a_{22} a_{14} a_{33} + a_{13} a_{23} a_{24} - a_{14} a_{23}^2\right)>0,
  \label{RN4a}
\end{aligned}
\end{equation}
where 
\begin{equation}
\begin{aligned}
& a_{11}=-\mu_1^2+3 \lambda_1 v_1^2+\frac{1}{2}\left(\bar{\lambda}_3 v_2^2+\bar{\lambda}_5 v_3^2+\bar{\lambda}_6 v_4^2+2 \lambda_{13} v_2 v_4\right), \\& a_{22}=-\mu_1^2+3 \lambda_1 v_2^2+\frac{1}{2}\left(\bar{\lambda}_3 v_1^2+\bar{\lambda}_5 v_4^2+\bar{\lambda}_6 v_3^2+2 \lambda_{13} v_1 v_3\right), \\
& a_{33}=-\mu_2^2+3 \lambda_2 v_3^2+\frac{1}{2}\left(\bar{\lambda}_4 v_4^2+\bar{\lambda}_5 v_1^2+\bar{\lambda}_6 v_2^2+2 \lambda_{14} v_1 v_4\right), \\&a_{44}=-\mu_2^2+3 \lambda_2 v_4^2+\frac{1}{2}\left(\bar{\lambda}_4 v_3^2+\bar{\lambda}_5 v_2^2+\bar{\lambda}_6 v_1^2+2 \lambda_{14} v_2 v_3\right), \\
& a_{12}=\bar{\lambda}_3 v_1 v_2+\frac{\bar{\lambda}_{11}}{2}  v_3 v_4+\lambda_{13}\left(v_1 v_4+v_2 v_3\right), \quad a_{13}=\bar{\lambda}_5 v_1 v_3+\lambda_{14} v_3 v_4+\frac{1}{2} \left(\bar{\lambda}_{11} v_2 v_4+ \lambda_{13} v_2^2\right), \\
& a_{14}=\bar{\lambda}_6 v_1 v_4+\lambda_{13} v_1 v_2+\frac{1}{2} \left(\bar{\lambda}_{11} v_2 v_3+\lambda_{14} v_3^2\right), \quad a_{23}=\bar{\lambda}_6 v_2 v_3+\lambda_{13} v_1 v_2+\frac{1}{2} \left(\bar{\lambda}_{11} v_1 v_4+ \lambda_{14} v_4^2\right), \\&
a_{24} = \bar{\lambda}_5 v_2 v_4 + \lambda_{14} v_3 v_4 + \frac{1}{2}\left( \bar{\lambda}_{11} v_1 v_3 + \lambda_{13} v_1^2 \right),
\quad a_{34}=\bar{\lambda}_4 v_3 v_4+\frac{\bar{\lambda}_{11}}{2}  v_1 v_2+\lambda_{14}\left(v_1 v_3+v_2 v_4\right) .
\end{aligned}
\end{equation}

For R-N-4b, the conditions for positive definiteness can be derived directly from those for R-N-4a by replacing the parameters $(v_1, v_2, v_3, v_4)$ with $(v_1, \pm v_2, \pm v_1, v_2)$. The case of R-N-4c can be treated similarly.

For R-I-1b, the conditions for positive definiteness are:
\begin{equation}
 \lambda_2>0,\quad \bar\lambda_{4}>2\lambda_{2},\quad \bar\lambda_{5}>\frac{2\mu_1^2}{v^2},\quad \bar\lambda_{6}>\frac{2\mu_1^2}{v^2}.
\end{equation}

For R-I-4a, the conditions for positive definiteness are identical to those for R-N-4a given in (\ref{RN4a}), except that the elements $a_{ij}$ therein are replaced by:
\begin{equation}
\begin{aligned}
& a_{11}=-\mu_1^2+3 \lambda_1 v_1^2+\frac{1}{2}\left(\bar{\lambda}_3 v_2^2+\bar{\lambda}_5 v_3^2\right) ,\quad a_{22}=-\mu_1^2+3 \lambda_1 v_2^2+\frac{1}{2}\left(\bar{\lambda}_3 v_1^2+\bar{\lambda}_6 v_3^2\right)+\lambda_{13} v_1 v_3, \\
& a_{33}=-\mu_2^2+3 \lambda_2 v_3^2+\frac{1}{2}\left(\bar{\lambda}_6 v_2^2+\bar{\lambda}_5 v_1^2\right), \quad a_{44}=-\mu_2^2+\frac{1}{2}\left(\bar{\lambda}_4 v_3^2+\bar{\lambda}_5 v_2^2+\bar{\lambda}_6 v_1^2+2 \lambda_{14} v_2 v_3\right) ,\\
& a_{12}=\bar{\lambda}_3 v_1 v_2+\lambda_{13} v_2 v_3, \quad a_{13}=\bar{\lambda}_5 v_1 v_3+\frac{\lambda_{13}}{2}  v_2^2, \quad a_{14}=\frac{1}{2}( \bar{\lambda}_{11} v_2 v_3+  \lambda_{14} v_3^2)+\lambda_{13} v_1 v_2,\\
& a_{23}=\bar{\lambda}_6 v_2 v_3+\lambda_{13} v_1 v_2, \quad a_{24}=\frac{1}{2}( \bar{\lambda}_{11} v_1 v_3+\lambda_{13} v_1^2), \quad a_{34}=\frac{\bar{\lambda}_{11}}{2}  v_1 v_2+\lambda_{14} v_1 v_3.
\end{aligned}
\end{equation}

For R-I-4b, the conditions for positive definiteness are identical to those for R-N-4a given in (\ref{RN4a}), except that the elements $a_{ij}$ are replaced by:
\begin{equation}
\begin{aligned}
& a_{11}=-\mu_1^2+3 \lambda_1 v_1^2+\frac{1}{2}(\bar{\lambda}_3 v_2^2+\bar{\lambda}_6 v_4^2+2 \lambda_{13} v_2 v_4), \quad a_{22}=-\mu_1^2+3 \lambda_1 v_2^2+\frac{1}{2}(\bar{\lambda}_3 v_1^2+\bar{\lambda}_5 v_4^2), \\
& a_{33}=-\mu_2^2+\frac{1}{2}(\bar{\lambda}_4 v_4^2+\bar{\lambda}_5 v_1^2+\bar{\lambda}_6 v_2^2+2 \lambda_{14} v_1 v_4), \quad a_{44}=-\mu_2^2+3 \lambda_2 v_4^2+\frac{1}{2}(\bar{\lambda}_5 v_2^2+\bar{\lambda}_6 v_1^2), \\
& a_{12}=\bar{\lambda}_3 v_1 v_2+\lambda_{13}v_1 v_4, \quad a_{13}=\frac{1}{2} (\bar{\lambda}_{11} v_2 v_4+\lambda_{13} v_2^2), \quad a_{14}=\bar{\lambda}_6 v_1 v_4+\lambda_{13} v_1 v_2, \\& a_{23}=\frac{1}{2} (\bar{\lambda}_{11} v_1 v_4+\lambda_{14} v_4^2)+\lambda_{13} v_1 v_2, \quad a_{24}=\bar{\lambda}_5 v_2 v_4+\frac{\lambda_{13}}{2}  v_1^2,\quad a_{34}=\frac{\bar{\lambda}_{11}}{2}  v_1 v_2+\lambda_{14}v_2 v_4 .
\end{aligned}
\end{equation}

For R-I-4c, the positive-definiteness conditions are identical to those for R-N-4a given in (\ref{RN4a}), except that the elements $a_{ij}$ therein are replaced by:
\begin{equation}
\begin{aligned}
& a_{11}=-\mu_1^2+3 \lambda_1 v_1^2+\frac{1}{2}(\bar{\lambda}_5 v_3^2+\bar{\lambda}_6 v_4^2), \quad a_{22}=-\mu_1^2+\frac{1}{2}(\bar{\lambda}_3 v_1^2+\bar{\lambda}_5 v_4^2+\bar{\lambda}_6 v_3^2+2 \lambda_{13} v_1 v_3), \\
& a_{33}=-\mu_2^2+3 \lambda_2 v_3^2+\frac{1}{2}(\bar{\lambda}_4 v_4^2+\bar{\lambda}_5 v_1^2+2 \lambda_{14} v_1 v_4), \quad a_{44}=-\mu_2^2+3 \lambda_2 v_4^2+\frac{1}{2}(\bar{\lambda}_4 v_3^2+\bar{\lambda}_6 v_1^2), \\
& a_{12}=\frac{\bar{\lambda}_{11}}{2}  v_3 v_4+\lambda_{13}v_1 v_4, \quad a_{13}=\bar{\lambda}_5 v_1 v_3+\lambda_{14} v_3 v_4, \quad a_{14}=\bar{\lambda}_6 v_1 v_4+\frac{\lambda_{14}}{2}  v_3^2, \\& a_{23}=\frac{1}{2} (\bar{\lambda}_{11} v_1 v_4+ \lambda_{14} v_4^2), \quad a_{24}=\frac{1}{2}( \bar{\lambda}_{11} v_1 v_3+ \lambda_{13} v_1^2)+\lambda_{14} v_3 v_4 ,\quad a_{34}=\bar{\lambda}_4 v_3 v_4+\lambda_{14}v_1 v_3 .
\end{aligned}
\end{equation}

For R-I-4d, the positive-definiteness conditions are identical to those for R-N-4a given in (\ref{RN4a}), except that the elements $a_{ij}$ therein are replaced by:
\begin{equation}
\begin{aligned}
&a_{11}=-\mu_1^2+\frac{1}{2}\left(\bar{\lambda}_3 v_2^2+\bar{\lambda}_5 v_3^2+\bar{\lambda}_6 v_4^2+2 \lambda_{13} v_2 v_4\right) ,\quad a_{22}=-\mu_1^2+3 \lambda_1 v_2^2+\frac{1}{2}\left(\bar{\lambda}_5 v_4^2+\bar{\lambda}_6 v_3^2\right), \\&
a_{33}=-\mu_2^2+3 \lambda_2 v_3^2+\frac{1}{2}\left(\bar{\lambda}_4 v_4^2+\bar{\lambda}_6 v_2^2\right),\quad a_{44}=-\mu_2^2+3 \lambda_2 v_4^2+\frac{1}{2}\left(\bar{\lambda}_4 v_3^2+\bar{\lambda}_5 v_2^2+2 \lambda_{14} v_2 v_3\right), \\&
a_{12}=\frac{\bar{\lambda}_{11}}{2}  v_3 v_4+\lambda_{13} v_2 v_3 ,\quad 
a_{13}=\frac{1}{2} (\bar{\lambda}_{11} v_2 v_4+ \lambda_{13} v_2^2)+\lambda_{14} v_3 v_4,\quad a_{14}=\frac{1}{2} (\bar{\lambda}_{11} v_2 v_3+ \lambda_{14} v_3^2) ,\\&
a_{23}=\bar{\lambda}_6 v_2 v_3+\frac{\lambda_{14}}{2}v_4^2,\quad a_{24}=\bar{\lambda}_5 v_2 v_4+\lambda_{14} v_3 v_4 ,\quad a_{34}=\bar{\lambda}_4 v_3 v_4+\lambda_{14} v_2 v_4.
\end{aligned}
\end{equation}

\subsection{Complex vacua}
\label{Hessiancomplex}

For complex vacua, positive-definiteness of the Hessian (\ref{Hessianc}) requires that all seven leading principal minors be strictly positive. Because of the complexity of the structure, we can explicitly provide only the first four conditions.

For C-N-2, the requirements that its first four leading principal minors of the Hessian be positive are identical to the four positive-definiteness conditions for R-N-2 given in (\ref{RN2}), except that the elements $a_{ij}$ therein are replaced by:
\begin{equation}
\begin{aligned}
& a_{11}=-\mu_1^2+3 \lambda_1 v_1^2+\frac{1}{2}\left[\bar{\lambda}_3 v_1^2+\left(\bar{\lambda}_5+\bar{\lambda}_6\right) v_3^2 \pm 2 \lambda_{13} v_{13}\right], \\
& a_{33}=-\mu_2^2+3 \lambda_2 v_3^2+\frac{1}{2}\left[\bar{\lambda}_4 v_3^2+\left(\bar{\lambda}_5+\bar{\lambda}_6\right) v_1^2 +(-1)^n2\lambda_{14} v_1 v_3\right], \\
& a_{12}= \pm \bar{\lambda}_3 v_1^2 \pm \frac{ \bar{\lambda}_{11}}{2}v_3^2+2 \lambda_{13} v_1 v_3, \quad a_{13}= \pm \left(\bar{\lambda}_5 + \frac{\bar{\lambda}_{11}}{2} \right)v_1 v_3+\frac{\lambda_{13}}{2} v_1^2\pm (-1)^n \lambda_{14} v_3^2, \\
& a_{14}=\left(\bar{\lambda}_6 +\frac{\bar{\lambda}_{11}}{2}\right)v_1v_3\pm \frac{\lambda_{13}}{2} v_1^2+(-1)^n\lambda_{14} v_3^2 ,\quad a_{34}= \pm\bar{\lambda}_4 v_3^2\pm\frac{\bar{\lambda}_{11}}{2}v_1^2\pm(-1)^n2\lambda_{14} v_1v_3 .
\end{aligned}
\end{equation}

For C-N-3a, the first four conditions for positive definiteness are identical to those for R-N-2 given in (\ref{RN2}), but the elements $a_{ij}$ therein are replaced by:
\begin{equation}
\begin{aligned}
& a_{11}=-\mu_1^2+3 \lambda_1 v_1^2+\frac{1}{2}\left[\bar{\lambda}_3 v_1^2+\left(\bar{\lambda}_5+\bar{\lambda}_6\right) v_3^2 \pm 2 \lambda_{13} v_1 v_3 \cos \left(\theta-\tfrac{n \pi}{5}\right)\right], \\
& a_{33}=-\mu_{2}^2+3 \lambda_2 v_3^2+\frac{1}{2}\left[\bar{\lambda}_4 v_3^2+\left(\bar{\lambda}_5+\bar{\lambda}_6\right) v_1^2+2 \lambda_{14} v_1 v_3 \cos \left(\theta+\tfrac{4 n \pi}{5}\right)\right], \\
& a_{12}= \pm \bar{\lambda}_3 v_1^2 \pm \frac{\bar{\lambda}_{11}}{2}v_3^2 \cos 2\left(\theta-\tfrac{n \pi}{5}\right)+2 \lambda_{13} v_1 v_3 \cos \left(\theta-\tfrac{n \pi}{5}\right), \\
& a_{13}= \pm \bar{\lambda}_5 v_1 v_3 \pm \frac{\bar{\lambda}_{11}}{2} v_1 v_3 \cos 2\left(\theta-\tfrac{n \pi}{5}\right)+\frac{\lambda_{13}}{2}v_1^2 \cos \left(\theta-\tfrac{n \pi}{5}\right) \pm \lambda_{14} v_3^2 \cos \left(\theta+\tfrac{4 n \pi}{5}\right), \\
& a_{14}=\bar{\lambda}_6 v_1 v_3+\frac{\bar{\lambda}_{11}}{2}v_1 v_3 \cos 2\left(\theta-\tfrac{n \pi}{5}\right) \pm \lambda_{13} v_1^2 \cos \left(\theta-\tfrac{n \pi}{5}\right)+\frac{\lambda_{14}}{2}v_3^2 \cos \left(\theta+\tfrac{4 n \pi}{5}\right),\\
& a_{34}= \pm\left[\bar{\lambda}_4 v_3^2+\frac{\bar{\lambda}_{11}}{2}v_1^2 \cos 2\left(\theta-\tfrac{n \pi}{5}\right)+2 \lambda_{14} v_1 v_3 \cos \left(\theta+\tfrac{4 n \pi}{5}\right)\right]. 
\end{aligned}
\end{equation}

For C-N-3b, the first four positive definiteness conditions are identical to those for R-N-2, given in (\ref{RN2}), except that the elements $a_{ij}$ are replaced by:
\begin{equation}
\label{HCN3b}
\begin{aligned}
& a_{11}=-\mu_1^2+3 \lambda_1 v_1^2+\frac{1}{2}\left[\bar{\lambda}_3 v_1^2+\left(\bar{\lambda}_5+\bar{\lambda}_6\right) v_3^2 \pm 2 \lambda_{13} v_1 v_3 \cos \theta\right], \\
& a_{33}=-\mu_2^2+3 \lambda_2 v_3^2+\frac{1}{2}\left[\bar{\lambda}_4 v_3^2+\left(\bar{\lambda}_5+\bar{\lambda}_6\right) v_1^2+2 \lambda_{14} v_1 v_3 \cos \theta\right] ,\\
& a_{12}= \pm \bar{\lambda}_3 v_1^2 \pm \frac{\bar{\lambda}_{11}}{2}v_3^2 \cos 2 \theta+2 \lambda_{13} v_1 v_3 \cos \theta, \\
& a_{13}= \pm \bar{\lambda}_5 v_1 v_3 \pm \frac{\bar{\lambda}_{11}}{2}v_1 v_3 \cos 2 \theta+\frac{\lambda_{13}}{2}v_1^2 \cos \theta \pm \lambda_{14} v_3^2 \cos \theta, \\
& a_{14}=\bar{\lambda}_6 v_1 v_3+\frac{\bar{\lambda}_{11}}{2}v_1 v_3 \cos 2 \theta \pm \lambda_{13} v_1^2 \cos \theta+\frac{\lambda_{14}}{2}v_3^2 \cos \theta ,\\
& a_{34}= \pm\left(\bar{\lambda}_4 v_3^2+\frac{\bar{\lambda}_{11}}{2} v_1^2 \cos 2 \theta+2 \lambda_{14} v_1 v_3 \cos \theta\right) .
\end{aligned}
\end{equation}

For C-N-3c, the first four conditions for positive definiteness are identical to those for R-N-2, as given in (\ref{RN2}), except that the elements $a_{ij}$ are replaced by:
\begin{equation}
\begin{aligned}
&a_{11}=-\mu_1^2+3 \lambda_1 v_1^2+\frac{1}{2}\left[\bar{\lambda}_3 v_1^2+\left(\bar{\lambda}_5+\bar{\lambda}_6\right) v_3^2\right]  ,\\&a_{33}=-\mu_2^2+3 \lambda_2 v_3^2+\frac{1}{2}\left[\bar{\lambda}_4 v_3^2+\left(\bar{\lambda}_5+\bar{\lambda}_6\right) v_1^2\right] ,\\
& a_{12}= \pm \bar{\lambda}_3 v_1^2 \mp \frac{\bar{\lambda}_{11}}{2}v_3^2 ,\quad  a_{13}= \left(\pm \bar{\lambda}_5  \mp \frac{\bar{\lambda}_{11}}{2} \right) v_1 v_3, \\& a_{14}=\left(\bar{\lambda}_6 -\frac{\bar{\lambda}_{11}}{2} \right) v_1 v_3,\quad a_{34}= \pm \bar{\lambda}_4 v_3^2 \mp \frac{\bar{\lambda}_{11}}{2}  v_1^2 .
\end{aligned}
\end{equation}

For C-N-3d, the positive definiteness conditions are identical to those for C-N-3c.

For C-N-4a, the first four positive definiteness conditions are identical to those for R-N-4a given in (\ref{RN4a}), except that the elements $a_{ij}$ are replaced by:
\begin{equation}
\begin{aligned}
&  a_{11}=-\mu_1^2+3 \lambda_1 v_1^2+\frac{1}{2}\left[\bar{\lambda}_3 v_2^2+\bar{\lambda}_5 v_3^2+\bar{\lambda}_6 v_4^2+2 \lambda_{13} v_2 v_4\right] ,\\
& a_{22}=-\mu_1^2+3 \lambda_1 v_2^2+\frac{1}{2}\left[\bar{\lambda}_3 v_1^2+\bar{\lambda}_5 v_4^2+\bar{\lambda}_6 v_3^2+2\lambda_{13} v_1 v_3\right], \\
& a_{33}=-\mu_2^2+3 \lambda_2 v_3^2+\frac{1}{2}\left[\bar{\lambda}_4 v_4^2+\bar{\lambda}_5 v_1^2+\bar{\lambda}_6 v_2^2+2 \lambda_{14} v_1 v_4\right], \\
& a_{44}=-\mu_2^2+3 \lambda_2 v_4^2+\frac{1}{2}\left[\bar{\lambda}_4 v_3^2+\bar{\lambda}_3 v_2^2+\bar{\lambda}_6 v_1^2+2 \lambda_{14} v_2 v_3\right] ,\\
& a_{12}=\bar{\lambda}_3 v_1 v_2+\frac{\bar{\lambda}_{11}}{2}  v_3 v_4+(-1)^n\lambda_{13}\left(v_1 v_4+v_2 v_3\right), \quad a_{13}=\bar{\lambda}_5 v_1 v_3+\frac{\bar{\lambda}_{11}}{2}  v_2 v_4+\frac{\lambda_{13}}{2}  v_2^2+\lambda_{14} v_3 v_4, \\
& a_{14}=\bar{\lambda}_6 v_1 v_4+\frac{\bar{\lambda}_{11}}{2}  v_2 v_3+(-1)^n\lambda_{13} v_1 v_2+\frac{\lambda_{14}}{2}  v_3^2 ,\quad a_{23}=\bar{\lambda}_6 v_2 v_3+\frac{\bar{\lambda}_{11}}{2}  v_1 v_4+(-1)^n \lambda_{13} v_1 v_2+\frac{\lambda_{14}}{2}  v_4^2 ,\\
& a_{24}=\bar{\lambda}_5 v_2 v_4+\frac{\bar{\lambda}_{11}}{2}  v_1 v_3+\frac{\lambda_{13}}{2}  v_1^2+\lambda_{14} v_3 v_4,\quad a_{34}=\bar{\lambda}_4 v_3 v_4+\frac{\bar{\lambda}_{11}}{2}  v_1 v_2+\lambda_{14}\left(v_1 v_3+v_2 v_4\right) .
\end{aligned}
\label{HRN4a}
\end{equation}

For C-N-4b, the positive-definiteness conditions follow directly from those for C-N-4a by replacing the parameters $(v_1, v_2, v_3, v_4)$ with $(v_1, \pm v_2, \pm v_1, v_2)$. The case of C-N-4c can be treated similarly.

For C-N-4d, the first four positive-definiteness conditions are identical to those for R-N-2 given in (\ref{RN2}), except that the elements $a_{ij}$ therein are replaced by:
\begin{equation}
\begin{aligned}
&  a_{11}=-\mu_1^2+3 \lambda_1 v_1^2+\frac{1}{2}\left[\bar{\lambda}_3 v_1^2+(\bar{\lambda}_5 +\bar{\lambda}_6 )v_3^2\pm2 \lambda_{13} v_1 v_3\right], \\
& a_{33}=-\mu_2^2+3 \lambda_2 v_3^2+\frac{1}{2}\left[\bar{\lambda}_4 v_3^2+(\bar{\lambda}_5 +\bar{\lambda}_6 )v_1^2+2 \lambda_{14} v_1 v_3\right] ,\\
& a_{12}=\pm\bar{\lambda}_3 v_1^2\pm\frac{\bar{\lambda}_{11}}{2}  v_3^2+(-1)^n2\lambda_{13}v_1v_3, \quad a_{13}=\pm\left(\bar{\lambda}_5+\frac{\bar{\lambda}_{11}}{2} \right) v_1 v_3+\frac{\lambda_{13}}{2}  v_1^2\pm\lambda_{14} v_3^2, \\
& a_{14}=\left(\bar{\lambda}_6 +\frac{\bar{\lambda}_{11}}{2} \right) v_1 v_3\pm(-1)^n\lambda_{13} v_1^2+\frac{\lambda_{14}}{2}  v_3^2 ,\quad a_{34}=\pm\bar{\lambda}_4 v_3^2\pm\frac{\bar{\lambda}_{11}}{2}  v_1^2\pm2\lambda_{14}v_1 v_3 .
\end{aligned}
\end{equation}

For C-I-4a, the first four conditions for positive definiteness are identical to those for R-N-4a given in (\ref{RN4a}), except that the elements $a_{ij}$ therein are replaced by:
\begin{equation}
\begin{aligned}
& a_{11}=-\mu_1^2+3 \lambda_1 v_1^2+\frac{1}{2}\left(\bar{\lambda}_3 v_2^2+\bar{\lambda}_5 v_3^2\right), \\& a_{22}=-\mu_1^2+3 \lambda_1 v_2^2+\frac{1}{2}\left(\bar{\lambda}_3 v_1^2+\bar{\lambda}_6 v_3^2+2 \lambda_{13} v_1 v_3\right), \\
& a_{33}=-\mu_2^2+3 \lambda_2 v_3^2+\frac{1}{2}\left(\bar{\lambda}_5 v_1^2+\bar{\lambda}_6 v_2^2\right) ,\\& a_{44}=-\mu_2^2+\frac{1}{2}\left(\bar{\lambda}_4 v_3^2+\bar{\lambda}_5 v_2^2+\bar{\lambda}_6 v_1^2+2 \lambda_{14} v_2 v_3 \cos \tfrac{n \pi}{5}\right), \\& a_{12}=\bar{\lambda}_3 v_1 v_2+\lambda_{13} v_2 v_3, \quad a_{13}=\bar{\lambda}_5 v_1 v_3+\frac{\lambda_{13}}{2}  v_2^2, \\&a_{14}=\left(\frac{\bar{\lambda}_{11}}{2}  v_2 v_3 +\lambda_{13} v_1 v_2\right) \cos \tfrac{3 n \pi}{5}+\frac{\lambda_{14}}{2}  v_3^2 \cos \tfrac{2 n \pi}{5}, \\& a_{23}=\bar{\lambda}_6 v_2 v_3+\lambda_{13} v_1 v_2, \quad a_{24}=\frac{1}{2}\left( \bar{\lambda}_{11} v_1 v_3 + \lambda_{13} v_1^2 \right)\cos \tfrac{3 n \pi}{5},\\
& a_{34}=\frac{\bar{\lambda}_{11}}{2} v_1 v_2 \cos \tfrac{3 n \pi}{5}+\lambda_{14} v_1 v_3 \cos \tfrac{2 n \pi}{5}.
\end{aligned}
\end{equation}

For C-I-4b, the first four conditions for positive definiteness are identical to those for R-N-4a given in (\ref{RN4a}), except that the elements $a_{ij}$ therein are replaced by:
\begin{equation}
\begin{aligned}
&a_{11}=-\mu_1^2+3 \lambda_1 v_1^2+\frac{1}{2}\left(\bar{\lambda}_3 v_2^2+\bar{\lambda}_6 v_4^2+2 \lambda_{13} v_2 v_4 \right), \quad a_{22}=-\mu_1^2+3 \lambda_1 v_2^2+\frac{1}{2}\left(\bar{\lambda}_3 v_1^2+\bar{\lambda}_5 v_4^2\right),\\
& a_{33}=-\mu_2^2+\frac{1}{2}\left(\bar{\lambda}_4 v_4^2+\bar{\lambda}_5 v_1^2+\bar{\lambda}_6 v_2^2+2 \lambda_{14} v_1 v_4 \cos \tfrac{n\pi}{5}\right), \quad a_{44}=-\mu_2^2+3 \lambda_2 v_4^2+\frac{1}{2}\left(\bar{\lambda}_5 v_2^2+\bar{\lambda}_6 v_1^2\right), \\
& a_{12}=\bar{\lambda}_3 v_1 v_2+\lambda_{13}v_1 v_4 , \quad a_{13}=\frac{1}{2}\left( \bar{\lambda}_{11} v_2 v_4+  \lambda_{13} v_2^2 \right)\cos \tfrac{3n\pi}{5}, \quad a_{14}=\bar{\lambda}_6 v_1 v_4+\lambda_{13} v_1 v_2, \\
& a_{23}=\frac{1}{2} \left(\bar{\lambda}_{11} v_1 v_4 +2\lambda_{13} v_1 v_2\right) \cos\tfrac{3n\pi}{5}+\frac{\lambda_{14}}{2} v_4^2 \cos  \tfrac{2n\pi}{5}, \quad a_{24}=\bar{\lambda}_5 v_2 v_4+\frac{\lambda_{13}}{2}  v_1^2,\\
& a_{34}=\frac{\bar{\lambda}_{11}}{2}  v_1 v_2 \cos  \tfrac{3n\pi}{5}+\lambda_{14} v_2 v_4 \cos\tfrac{2n\pi}{5}.
\end{aligned}
\end{equation}

For C-I-4c, the first four conditions for positive definiteness are identical to those for R-N-4a given in (\ref{RN4a}), except that the elements $a_{ij}$ are replaced by:
\begin{equation}
\begin{aligned}
&a_{11}=-\mu_1^2+3 \lambda_1 v_1^2+\frac{1}{2}\left(\bar{\lambda}_5 v_3^2+\bar{\lambda}_6 v_4^2\right), \quad a_{22}=-\mu_1^2+\frac{1}{2}\left(\bar{\lambda}_3 v_1^2+\bar{\lambda}_5 v_4^2+\bar{\lambda}_6 v_3^2+2 \lambda_{13} v_1 v_3 \cos \tfrac{3n\pi}{5}\right), \\
& a_{33}=-\mu_2^2+3 \lambda_2 v_3^2+\frac{1}{2}\left(\bar{\lambda}_4 v_4^2+\bar{\lambda}_5 v_1^2+2 \lambda_{14} v_1 v_4 \right), \quad a_{44}=-\mu_2^2+3 \lambda_2 v_4^2+\frac{1}{2}\left(\bar{\lambda}_4 v_3^2+\bar{\lambda}_6 v_1^2\right), \\
& a_{12}=\frac{\bar{\lambda}_{11}}{2}  v_3 v_4 \cos \tfrac{n\pi}{5}+\lambda_{13} v_1 v_4 \cos \tfrac{4n\pi}{5}, \quad a_{13}=\bar{\lambda}_5 v_1 v_3+\lambda_{14} v_3 v_4 ,\\ &a_{14}=\bar{\lambda}_6 v_1 v_4+\frac{\lambda_{14}}{2}  v_3^2, \quad a_{23}=\frac{1}{2}( \bar{\lambda}_{11} v_1 v_4 + \lambda_{14} v_4^2 )\cos \tfrac{n\pi}{5},\\
& a_{24}=\frac{1}{2} (\bar{\lambda}_{11} v_1 v_3+2\lambda_{14} v_3 v_4) \cos \tfrac{n\pi}{5}+\frac{\lambda_{13}}{2} v_1^2 \cos \tfrac{4n\pi}{5},\quad a_{34}=\bar{\lambda}_4 v_3 v_4+\lambda_{14} v_1 v_3 . 
\end{aligned}
\end{equation}

For C-I-4d, the first four conditions for positive definiteness are identical to those for R-N-4a given in (\ref{RN4a}), except that the elements $a_{ij}$ are replaced by:
\begin{equation}
\begin{aligned}
& a_{11}=-\mu_1^2+\frac{1}{2}\left(\bar{\lambda}_3 v_2^2+\bar{\lambda}_5 v_3^2+\bar{\lambda}_6 v_4^2+2 \lambda_{13} v_2 v_4 \cos \tfrac{n \pi}{5}\right) ,\quad a_{22}=-\mu_1^2+3 \lambda_1 v_2^2+\frac{1}{2}\left(\bar{\lambda}_5 v_4^2+\bar{\lambda}_6 v_3^2\right), \\
& a_{33}=-\mu_2^2+3 \lambda_2 v_3^2+\frac{1}{2}\left(\bar{\lambda}_4 v_4^2+\bar{\lambda}_6 v_2^2\right), \quad a_{44}=-\mu_2^2+3 \lambda_2 v_4^2+\frac{1}{2}\left(\bar{\lambda}_4 v_3^2+\bar{\lambda}_5 v_2^2+2 \lambda_{14} v_2 v_3\right) ,\\
& a_{12}=\frac{\bar{\lambda}_{11}}{2}  v_3 v_4 \cos \tfrac{2 n \pi}{5}+\lambda_{13} v_2 v_3 \cos \tfrac{3 n \pi}{5} ,\quad a_{13}=\left(\frac{\bar{\lambda}_{11}}{2}  v_2v_4+\lambda_{14} v_3v_4\right)  \cos \tfrac{2 n \pi}{5}+\frac{\lambda_{13} }{2} v_2^2 \cos \tfrac{3 n \pi}{5}, \\
& a_{14}=\frac{1}{2}\left( \bar{\lambda}_{11} v_2v_3+\lambda_{14} v_3^2\right)  \cos \tfrac{2 n \pi}{5} ,\quad a_{23}=\bar{\lambda}_6 v_2 v_3+\frac{\lambda_{14}}{2}  v_4^2, \\
& a_{24}=\bar{\lambda}_5 v_2v_4+\lambda_{14} v_3 v_4,\quad a_{34}=\bar{\lambda}_4 v_3 v_4+\lambda_{14} v_2 v_4 .
\end{aligned}
\end{equation}

\end{document}